\documentclass[10pt]{article}

\usepackage{amsmath}
\usepackage{amssymb}

\usepackage{graphicx}
\usepackage{caption}

\usepackage{cite}

\usepackage{color}

\usepackage{float}

\usepackage[labelfont=bf,labelsep=period,justification=raggedright]{caption}

\makeatletter
\renewcommand{\@biblabel}[1]{\quad#1.}
\makeatother

\date{}

\begin{document}

\begin{flushleft}
{\Large
\textbf{The effect of ionic diffusion on extracellular potentials in neural tissue}
}
\\
Geir Halnes$^{1,\ast}$,
Tuomo M\"{a}ki-Marttunen$^{2}$
Daniel Keller$^{3}$,
Klas H. Pettersen$^{4}$
Gaute T. Einevoll$^{1}$
\\
\bf{1} Dept. of Mathematical Sciences and Technology, Norwegian University of Life Sciences, {\AA}s, Norway
\\
\bf{2} NORMENT, KG Jebsen Centre for Psychosis Research, Institute of Clinical Medicine, University of Oslo, Oslo, Norway
\\
\bf{3} Blue Brain Project, École Polytechnique Fédérale de Lausanne (EPFL), Lausanne, Switzerland
\\
\bf{4} Letten Centre and GliaLab, Institute of Basic Medical Sciences, University of Oslo, Oslo, Norway
\\
$\ast$ E-mail: geir.halnes@nmbu.no
\end{flushleft}


\section*{Abstract}
In computational neuroscience, it is common to use the simplifying assumption that diffusive currents are negligible compared to Ohmic currents. However, endured periods of intense neural signaling may cause local ion concentration changes in the millimolar range. Theoretical studies have identified scenarios where steep concentration gradients give rise to diffusive currents that are of comparable magnitude with Ohmic currents, and where the simplifying assumption that diffusion can be neglected does not hold. We here propose a novel formalism for computing (1) the ion concentration dynamics and (2) the electrical potential in the extracellular space surrounding multi-compartmental neuron models or networks of such (e.g., the Blue-Brain simulator). We use this formalism to explore the effects that diffusive currents can have on the extracellular (ECS) potential surrounding a small population of active cortical neurons. Our key findings are: (i) Sustained periods of neuronal output (simulations were run for 84 s) could change local ECS ion concentrations by several mM, as observed experimentally. (ii) For large, but realistic, concentration gradients, diffusive currents in the ECS were of the same magnitude as Ohmic currents. (iii) Neuronal current sources could induce local changes in the ECS potential by a few mV, whereas diffusive currents could could induce local changes in the ECS potential by a few tens of a mV. Diffusive currents could thus have a quite significant impact on ECS potentials. (v) Potential variations caused by diffusive currents were quite slow, but could influence the comparable to those induced by Ohmic currents up to frequencies as high as 7Hz.

\section*{Introduction}
During periods of intense neural signaling, ion concentrations in the extracellular space (ECS) can change locally by several mM \cite{Cordingley1978, Dietzel1989, Gardner-Medwin1983, Chen2000, Haj-Yasein2014}. For example, the the extracellular K\textsuperscript{+}-concentration can increase from a typical basal level of around 3mM and up to levels between 8 and 12 mM during non-pathological conditions \cite{Hertz2013, Chen2000, Newman1993}. Ion concentration shifts in the ECS will change neuronal reversal potentials and firing patterns (see e.g., \cite{Oyehaug2012, Park2006}). Too large deviances from basal levels can lead to pathological conditions such as hypoxia, anoxia, ischemia and spreading depression \cite{Sykova2008}. The extracellular ion concentration dynamics depends not only on transmembrane ionic sources, but also on ionic diffusion along extracellular concentration gradients. Diffusion plays an important role in maintaining local ion concentrations at healthy levels \cite{Nicholson2000}, but may also be involved in the propagation of e.g. epileptic seizures during pathological conditions \cite{Florence2009}.

Diffusion of ions along concentration gradients carry electrical currents \cite{Kofuji2004, Halnes2013}, and may in principle have measurable effects on electrical potentials (c.f. liquid junction potentials \cite{VanEgeraat1993, Sokalski2001, Perram2006}). In many theoretical approaches, it is assumed that diffusive currents are negligible compared to the Ohmic currents propelled by the electrical field (hereby termed field currents). This is, for example, an underlying assumption when intracellular currents are computed with the cable equation, which is used in most multi-compartmental neural models (see e.g., \cite{Rall1977}). It is also an underlying assumption in volume conductor theory, which is used to estimate extracellular potentials \cite{Holt1999, Pettersen2008, Pettersen2008a, Reimann2013}, and in standard current source density (CSD) theory, which predicts neural current sources from recordings of extracellular potentials (see e.g., \cite{Einevoll2007, Pettersen2008a}).

Ion concentration changes in the ECS often are accompanied by a slow negative potential shift, which can be in the order of a few millivolts \cite{Sykova2008, Kriz1975, Lothman1975, Dietzel1989, Cordingley1978}. Although Dietzel et al. argued that glial potassium buffering currents were the main source of these slow potential shifts, they also discussed possible contributions from diffusive currents \cite{Dietzel1989}. Theoretical studies have also identified scenarios where large, but biologically realistic ion concentration gradients may induce diffusive currents that are comparable in magnitude to Ohmic currents, both in the intracellular \cite{Qian1989} and extracellular \cite{Halnes2013} space. As ion concentrations in the ECS typically change at the time scale of seconds \cite{Dietzel1989, Chen2000, Halnes2013}, it is unclear whether diffusive currents that originate from such changes can influence recorded extracellular potentials at frequencies higher than the typical cut-off frequencies of 0.1-0.2Hz applied in most electrode systems (see e.g., \cite{Einevoll2007}). However, if diffusive currents can have an impact on recorded extracellular potentials, it calls for a re-interpretation of extracellular voltage recordings, which have typically been assumed to predominantly reflect transmembrane current sources (and i.e. not diffusion in the ECS). This was the topic of a recent debate, where the omittance of diffusive currents in CSD theory was suggested as a possible explanation to an observed discrepancy between theoretical predictions and empirical measurements \cite{Riera2012, Riera2013, Destexhe2012, Gratiy2013, Cabo2014}.

In principle, one could combine experimental measurements of ion concentration gradients with prior knowledge of diffusion constants and typical tortuosities (hindrances) \cite{Nicholson1981, Nicholson1998} to determine the magnitude of the diffusive currents in the ECS. However, net diffusive currents generally depend on the joint movement of multiple ion species along their respective concentration gradients \cite{Halnes2013}. For example, in regions with intense neural action potential (AP) firing, we would expect the extracellular [K\textsuperscript{+}] and [Na\textsuperscript{+}] to increase and decrease, respectively. Accordingly, we would expect K\textsuperscript{+} to diffuse out from such a region and a similar amount of Na\textsuperscript{+} to diffuse into the same region. The net transport of charge (diffusive current) into/out from such a region would thus be much smaller than the charge transported by K\textsuperscript{+} and Na\textsuperscript{+} separately. A reliable experimental estimate of the net diffusive current would therefore require an extreme accuracy in the measurement of the different concentration gradients, and is probably not feasible. Therefore, questions about the relative importance of extracellular diffusive currents may be better addressed by a theoretical electrodiffusive framework. To give good predictions, such a framework should ensure both biophysical realism and physical consistency, and should ensure (i) that extracellular ion concentration gradients have values that are realistic in neural tissue, (ii) that ion concentration gradients of different species are consistent in the sense that they ensure local electroneutrality in the bulk solution, (iii) that realistic extracellular electrical potentials consistently follows from the interplay between neural, transmembrane current sources, and extracellular diffusive sources.

Several modelling frameworks based on the Poisson-Nernst-Planck equations are available for for handling the time development of the ion concentration dynamics and electrical potential \cite{Leonetti1998, Lu2007, Lopreore2008, Nanninga2008, Pods2013}. These frameworks typically require a very fine spatial and temporal resolution, and are not well suited for predictions at a population level. Simplified models, on the other hand, have so far either been restricted to only considering the ICS \cite{Qian1989}, or to be limited in the handling of neuronal features, such as ion channel distributions and morphology \cite{Halnes2013}. The standard tool for simulating morphologically complex neurons is the NEURON simulator, which is an effective tool for computing the dynamics of neuronal membrane potentials. There are methods toolboxes for computing extracellular fields surrounding neural models based on the NEURON simulator (e.g., \cite{Linden2014}), but no formalism has yet been developed to compute the extracellular ion concentration dynamics and diffusion in a way that can readily be combined with NEURON.

\begin{figure}[!ht]
\begin{center}
\includegraphics[width=6in]{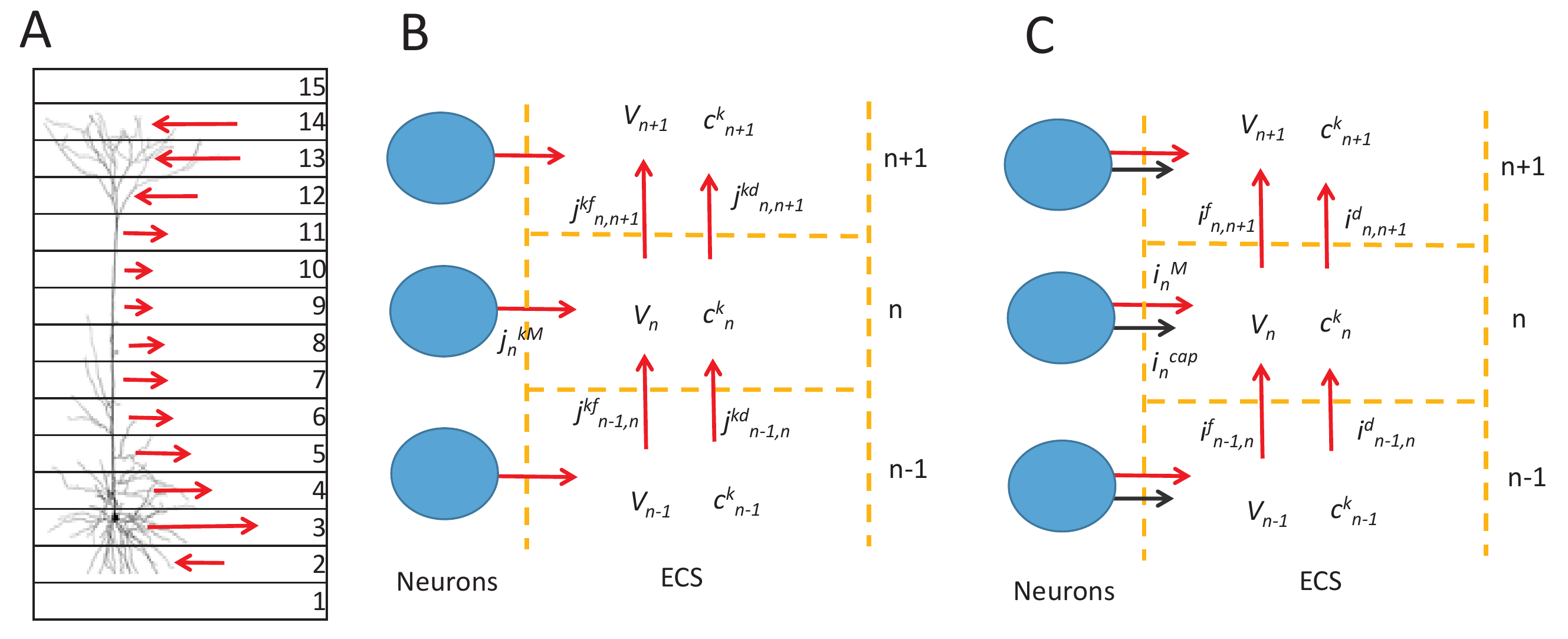}
\end{center}
\caption{\textbf{The model system.} \textbf{(A)} A cortical column was subdivided into 15 depth intervals. The edges $n=1$ and $n=15$ were assumed to be a constant background. The interior 13 depth intervals contained a population of 10 neurons (only one shown in Figure) simulated with the NEURON simulator. The output of specific ions into each depth interval $n$ was computed for 10 neurons and summed, yielding the total input of an ion species $k$ to each depth interval. \textbf{(B)} Ion concentration dynamics in a ECS subvolume $n$. $J_n^{kM}$ denotes the total transmembrane efflux of ion species $k$ into the depth interval $n$ of the whole population of neurons. $J^{kf}$ and $J^{kd}$ denote ECS fluxes between neighboring subvolume driven by electrical potential differences and diffusion, respectively. \textbf{(C)} The extracellular potential could be calculated by demanding that the sum of currents into each ECS subvolume should be zero. Currents were determined by summing the contributions from all ionic fluxes, and adding the capacitive current (black arrow).
}
\label{Fvox}
\end{figure}

In the current work, we have explored the possible effects that diffusive currents may have on the extracellular potential around a small population of cortical neurons. To do this, we developed a hybrid formalism which is briefly summarized in Fig. \ref{Fvox}. To simulate the neurodynamics, we used a previously developed multi-compartmental neural model of layer 5 pyramidal neurons with realistic morphology which was implemented in the NEURON simulator \cite{Hay2011}), and driven by realistic synaptic input (Fig. \ref{Fvox}A). When we had computed the neuronal output in this way, another, electrodiffusive framework was used for computing the dynamics of the ion concentrations and electrical potential in the ECS surrounding the neurons (Fig. \ref{Fvox}B). The scheme was based on the Nernst-Planck equations for electrodiffusion. However, instead of deriving the extracellular potential from Poisson's equation, we derived it from Kirchoff's current law (demanding that all currents into a ECS subvolume should sum to zero), a physical constraint that is valid at a larger spatiotemporal resolution (Fig. \ref{Fvox}C). This framework represents a novel, efficient and generally applicable method for simulating extracellular dynamics surrounding multi-compartmental neural models or networks of such (e.g., the Blue Brain simulator).

This work is organized in the following way: In the result-section, the main focus is put on the investigation of the possible role of diffusive currents on electrical potentials in the ECS. In the Discussion section, we discuss possible implications that our findings will have for the interpretation of data from extracellular recordings. The Discussion also includes an overview of the assumptions made in the presented model, and on how the framework can be expanded to allow for more thorough investigations of extracellular diffusive currents in neural tissue. A detailed derivation of the electrodiffusive formalism is found in the Methods-section (which is found at the end of in this article).

\section*{Results}
We investigated the role that diffusion can have on extracellular potentials by comparing simulations where the ECS dynamics was predicted using the electrodiffusive formalism with simulations where diffusive currents were assumed to be negligible, so that extracellular ion transports were solely due to field currents. The model simulation upset is briefly introduced in the following section and in Fig. \ref{Fvox}, while the details are postponed to the methods-section. A list of useful symbols and definitions can be found in Table \ref{T1}.

\begin{table}[!ht]
\caption{\textbf{Useful symbols and definitions}}
\begin{tabular}{|l|l|l|}
  \textbf{Symbol} & \textbf{Explanation} & \textbf{Units} \\
  $J^{kM}_n$ & Net membrane flux of species $k$ into subvolume $n$ & mol/s \\
  $I^M_n$ & Net ionic membrane current into subvolume $n$ & A \\
  $I^{cap}_n$ & Capacitive current into subvolume $n$ & A \\
  $V_n$ & Extracellular potential in subvolume $n$ & V \\
  $J^{kf}_{n-1,n}$ & Flux of $k$ from subvolume $n-1$ to $n$ due to electrical migration & mol/s\\
  $I^{f}_{n-1,n}$ & Electrical field current from subvolume $n-1$ to $n$  & A\\
  $J^{kd}_{n-1,n}$ & Flux of $k$ from subvolume $n-1$ to $n$ due to diffusion & mol/s \\
  $I^{d}_{n-1,n}$ & Diffusive current from subvolume $n-1$ to $n$ &  A \\
  $l_{c}$ & Length of ECS subvolume edge & $100 \mu m$ \\
  $A_{c}$ & Cross section area of ECS subvolume & $600 \mu m^2$ \\
  \end{tabular}
  \noindent
\begin{flushleft}
\end{flushleft}
\label{T1}
\end{table}

\subsection*{Dynamics of a small neuronal population}
Since diffusion takes place at a long time scale compared to the millisecond time scale of neuronal firing, we simulated our neural population for 84 s.  The neurons were driven by Poissonian input trains through 1000 synapses distributed uniformly over their membrane. The 10 neurons were simulated independently, receiving different input trains (but with the same statistics). Synaptic weights were tuned so that the input evoked an average single-neuron AP firing rate of about 5 APs per second, which is a typical firing rate for cortical neurons \cite{Linden2011}.

\begin{figure}[!ht]
\centering
\includegraphics[width=6in]{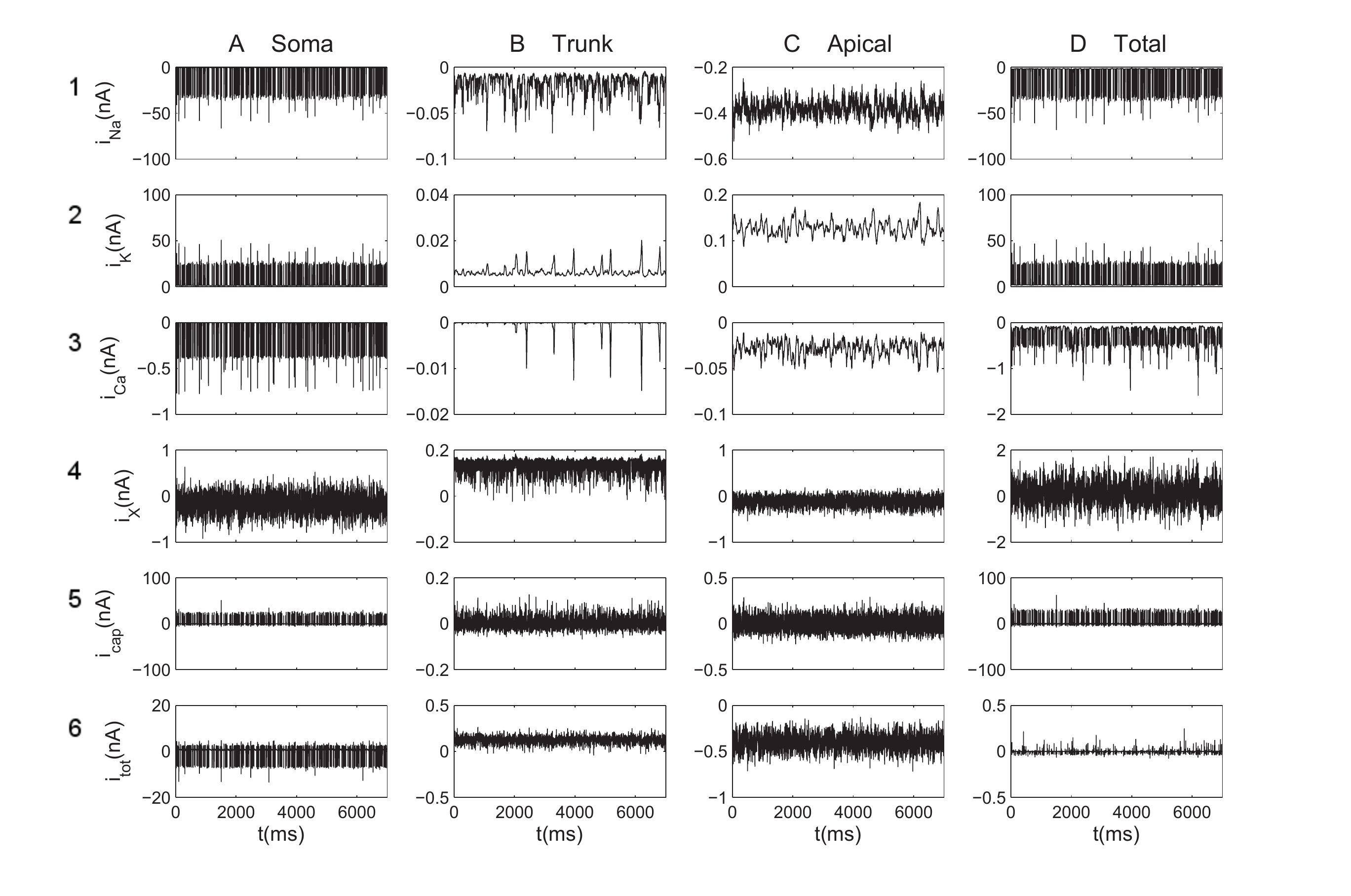}
\caption{\textbf{Output from the neuronal population.} Transmembrane currents into selected extracellular volumes, including \textbf{(A)} the region containing  the neuronal somatas ($n=3$), \textbf({(B)} the region containing the trunk of the apical dendrite ($n=7$), and \textbf{(C)} the region where the apical dendrites branched out ($n=13$). Currents were subdivided into ion specific currents (row 1-4) and the capacitive current (row 5). The sum of all currents into a subvolume $n$ is indicated in row 6. The transmembrane currents were defined as positive when crossing the membrane in the outward direction. Currents were summed over all neural segments (of all neurons) that occupied a given ECS-volume (depth level $n$). The total transmembrane currents of the neuron as a whole (summed over all $N-2$ depth levels) were also calculated \textbf{(D)}.
}
\label{Fdata}
\end{figure}

The time-development of the output of the neural population to three selected extracellular volumes is shown in Fig. \ref{Fdata} for the first 7s of the simulation. Fig. \ref{Fdata}A shows the currents into the the ECS volume containing the somatas. Here, we see clear signatures of neuronal APs: Neuronal depolarization by a brief Na\textsuperscript{+} current pulses (of about -50nA, since this current is leaving the ECS), and repolarization by a brief K\textsuperscript{+} current (of about +50nA, since this current enters the ECS). Generally, the subvolume containing the somatas received a significantly higher influx/efflux of ions (Fig. \ref{Fdata}A) compared to the volumes containing the apical trunk (Fig. \ref{Fdata}B) and branches (Fig. \ref{Fdata}C). These differences were mainly because the soma region contained a higher neuronal membrane area than other subvolumes. Similarly, the region where the apical dendrite branches, contained a larger membrane area than a part of the apical dendritic trunk, so that the net transmembrane currents were bigger in Fig. \ref{Fdata}C compared to Fig. \ref{Fdata}B. Regional differences were also due to the distribution of specific ion channels in the neuronal model \cite{Hay2011}. For example, in the region containing the somata and initial parts of axons/dendrites, the density of Na\textsuperscript{+} and K\textsuperscript{+}-channels was much larger than in regions containing only dendrites. Accordingly, the Na\textsuperscript{+} and K\textsuperscript{+} currents leaving/entering the ECS in the soma-region (Fig. \ref{Fdata}A1-2) were almost identical to the total Na\textsuperscript{+} and K\textsuperscript{+} currents that left/entered the neuron as a whole (Fig. \ref{Fdata}D1-2).

The neural output in Fig. \ref{Fdata} (but for 84 s, and not only the 7 s shown in the figure) was used in all simulations shown below.

\subsubsection*{The effect of diffusion on slow extracellular potentials}
\begin{figure}[!ht]
\begin{center}
\includegraphics[width=6in]{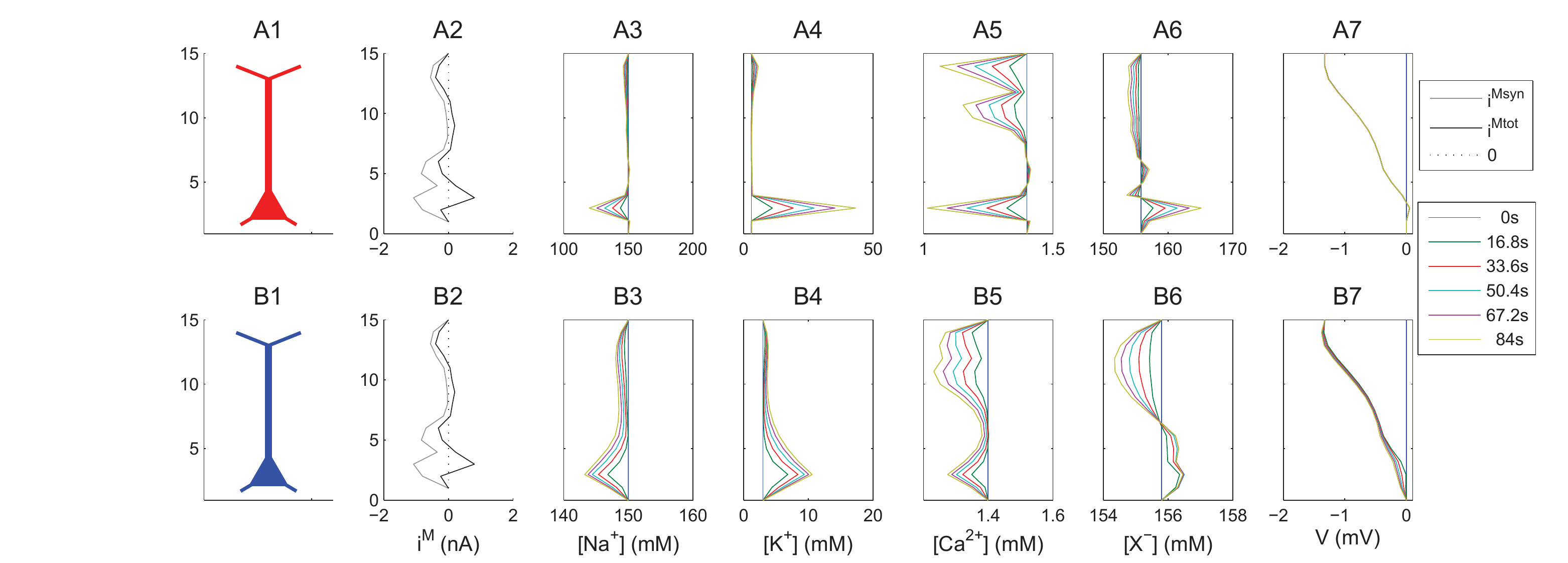}
\end{center}
\caption{\textbf{Extracellular profiles in ion concentrations and potential.} Distribution of different system variables over the depth of the cortex for the situation where diffusion was assumed to be zero \textbf{(A)}, and for the situation with diffusion included \textbf{(B)}. The distribution of transmembrane current sources (temporal average) is shown in (A2) and (B2). Panels A3-A6 and B3-B6 show the distributions of ECS ion concentrations at different time points. Panels A7 and B7 show the low pass filtered ECS potential ($\langle V \rangle$). For $\langle V \rangle$ (A7, B7), the temporal average was taken over the time interval 16.8s prior to the value indicated in the legend. For the current sources (A2,B2), the temporal average was taken over the first 16.8 ms of the simulation (the statistics of current sources did not change during the simulation, and any time interval $>1s$ gave nearly identical results for the temporal average).
}
\label{Fconc}
\end{figure}

Extracellular potentials may have many origins, including synaptic currents, passive neuronal membrane currents, slow or fast active neuronal membrane currents \cite{Pettersen2008a, Reimann2013}, or glial buffering currents \cite{Dietzel1989}. To explore whether also diffusive currents in the ECS can be sources that influence the ECS potential, we first studied the long time-scale dynamics of the system. As we saw in Fig.\ref{Fdata}, the local neuronal current sources varied on a fast temporal time scale. However, the statistics of the input remained the same throughout the 84 s simulation, and on a time resolution of seconds, the transmembrane current source (at a given location, $n$) remained roughly constant with time. Fig.\ref{Fconc}A2 and B2 show the profiles of the transmembrane current sources over the depth of the cortex when averaged over a time interval of 7 s. More or less identical profiles were observed if the average was taken over another time interval, regardless of where in the 84 s simulation this interval was placed, or the length of the interval as long as it exceeded a few seconds (results not shown). In the soma-region ($n=3$) a net current entered the ECS (current source), whereas a net current left the ECS in all other subvolumes (black lines in Fig. \ref{Fconc}A2 and B2). This observation indicates that there should be a net extracellular current in the positive $z$-direction (in order to complete the current loop).

As the neurodynamics resulted in influxes or effluxes of different ion species, ionic concentrations in the ECS varied during the 84 s simulations. Fig.\ref{Fconc} shows the extracellular concentration profiles at selected time points in the cases when (i) diffusion was not included (Fig.\ref{Fconc}A3-6), and (ii) when diffusion \emph{was} included (Fig.\ref{Fconc}B3-6). In the case (i) with no diffusion, changes in ECS ion concentrations were sharp in the soma region ($n=3$), as the transmembrane sources were larger there. For example, the Na\textsuperscript{+} concentration decreased significantly, while K\textsuperscript{+} increased significantly in the soma region. This effect was less pronounced in the case (ii) when diffusion was included, as diffusion acted to smoothen the ion concentrations over the cortical depth.

In the case (ii) when diffusion was included, the K\textsuperscript{+}-concentration in the soma region increased from a basal level of 3 mM to slightly above 10 mM during the simulation. These final ECS concentrations were in line with the maximum local changes that have been observed experimentally during non-pathological conditions \cite{Hertz2013, Chen2000, Newman1993}. Hence, we expect that the extracellular diffusive transports obtained with such ion concentration gradients are realistic. In the case (i) when diffusion was not included, the local ion concentration changes in the soma-regions were unrealistically high. However, this was of no concern for our simulation of other variables, as no system properties depended on the ECS ion concentrations in the case when diffusion was not included.

To explore whether diffusion could have an impact on ECS potentials, we calculated the low pass filtered ECS potential ($\langle V \rangle$) in the cases without (i) and with (ii) diffusion included (Fig. \ref{Fconc}A7 and B7). Our measure for ($\langle V \rangle$) at a time $t$ was simply the average $V$ taken over the time interval between two of the selected time points in Fig. \ref{Fconc} (i.e. $\langle V \rangle$ at time $t$ was the temporal average of $V$ between $t-16.8 s$ and $t$). As the figure shows, $\langle V \rangle$ varied from around 0 mV in the basal dendrites and soma-region, to about -2 mV in the apical dendrites. Basically the profile of $\langle V \rangle$ reflected the profile of the transmembrane current (Fig. \ref{Fconc}A2 and B2). As a net current entered the ECS in the soma region, and left the ECS along the apical dendrites, an extracellular current in the positive $z$-direction was required in order to \emph{close the loop}. A positive current in the $z$-direction, is consistent with a negative voltage gradient in the positive $z$-direction. The $\langle V \rangle$ profile obtained in our simulations were qualitatively similar to what have been seen experimentally, where profiles of sustained voltages which vary by a few mV across the depth of the cortex have been observed (see e.g. \cite{Dietzel1989}).

When diffusion was not included, the $\langle V \rangle$ profiles did not vary with time (profiles for different $t$ coincide in Fig. \ref{Fconc}A7 shows). This was as expected, since $\langle V \rangle$ in this case was determined solely by transmembrane current sources, which, as we saw in Fig. \ref{Fconc}A2 were constant over time (for a time resolution of seconds). The $\langle V \rangle$ profiles obtained when diffusion was included were different (Fig. \ref{Fconc}B7). In the latter case, $\langle V \rangle$ did vary with time, especially in the soma region, where the local ECS potential was reduced by approximately 0.2 mV during the time course of the simulation. Thus, Fig. \ref{Fconc}B7 shows that diffusion can indeed induce visible changes in the ECS potential. Several previous, experimental studies have observed slow negative potential shift measured in the ECS that coincide with changes in ion concentrations, and in particular increases in extracellular K\textsuperscript{+} \cite{Kriz1975, Lothman1975, Cordingley1978, Dietzel1989}.

\subsection*{Mechanisms behind the impact of diffusion on extracellular potentials}
\begin{figure}[!ht]
\begin{center}
\includegraphics[width=6in]{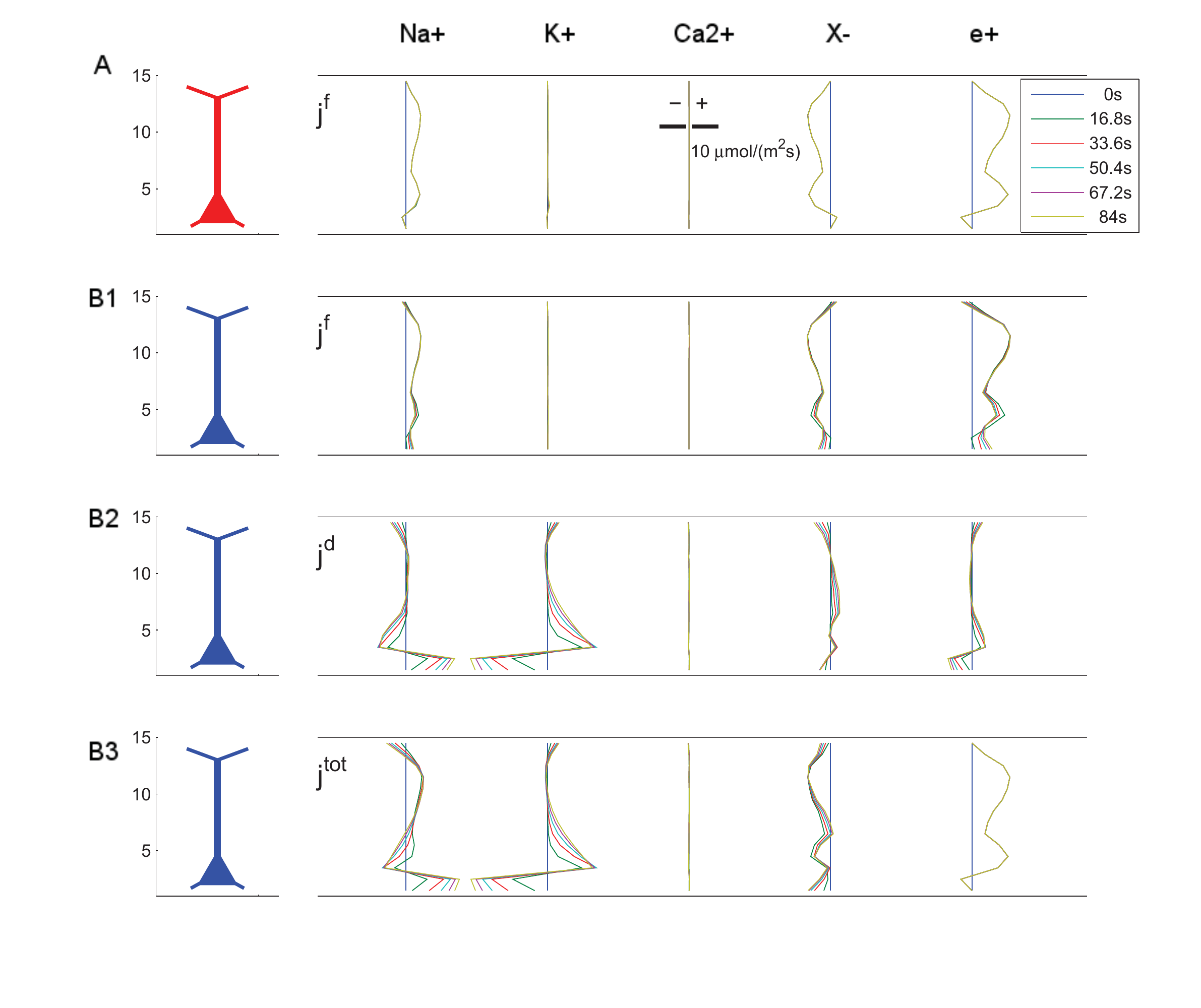}
\end{center}
\caption{\textbf{Extracellular Ion fluxes.} Low pass filtered extracellular fluxes in the case when diffusion was assumed to be zero \textbf{(A)}, and in the case when diffusion was included \textbf{(B)}. In the latter case, the total flux \textbf{(B3)} was subdivided into a field-driven \textbf{(B1)} and diffusive \textbf{(B2)} component. Curves to the right/left of the blue, vertical lines represent fluxes in the positive/negative $z$-direction respectively. Electrical currents were represented as fluxes of positive unit charges $j^{e+}$ = $i/F$ (rightmost column). The scale bar and legend apply to all fluxes. The low pass filtered fluxes were computed as the temporal mean of the real fluxes, taken over the 16.8s prior to the value indicated in the legend.
}
\label{Fflows}
\end{figure}

To explore the relationship between diffusion and $<V>$ in further detail, we plotted the extracellular fluxes of all ion species (K\textsuperscript{+}, Na\textsuperscript{+},  Ca\textsuperscript{2+}, and X\textsuperscript{-}) in the cases (i) without and (ii) with diffusion included at the selected time points (Fig. \ref{Fflows}). We also plotted the net electrical current associated with the ionic fluxes. To be able to compare the electrical current directly with ionic fluxes, we represented the currents as a flux of positive unit charges $e^+$. When diffusion was not included, all transports in the ECS were driven by the electrical field (Fig. \ref{Fflows}A). The main transports were those mediated by the most abundant ion species in the ECS, which in our simulation were Na\textsuperscript{+} and X\textsuperscript{-}. Due to the negative potential gradient between the soma and apical dendrites (\ref{Fconc}A7), Na\textsuperscript{+} was driven away from the soma, while X\textsuperscript{-} (having the opposite valence) was driven towards the soma. Both these transport amounted to a net electrical current away from the soma, i.e. the flux of positive unit charges was positive in depth intervals with $n>3$, and negative in depth intervals with $n<3$.

When diffusion was included, we distinguished between extracellular transports due the electrical field (Fig. \ref{Fflows}B1), and extracellular transports due to diffusion (Fig. \ref{Fflows}B2). The transports due to the electrical field were qualitatively similar to what we saw in the case without diffusion (compare Fig. \ref{Fflows}A with Fig. \ref{Fflows}B1)). However, in the case when diffusion was included, these transports varied slightly with time, since the extracellular $\langle V \rangle$ profile varied with time (as we saw in Fig. \ref{Fconc}B7). Unlike the field currents, the diffusive transports were not determined by which ion species that were most abundant in the ECS, but on the concentration gradients, which were largest for Na\textsuperscript{+} and K\textsuperscript{+}. Due to the high decrease/increase of Na\textsuperscript{+}/K\textsuperscript{+} caused by AP firing in the soma region, extracellular diffusion drove Na\textsuperscript{+} into the soma region, and K\textsuperscript{+} out from the soma region (Fig. \ref{Fflows}B2). Due to the opposite directionality of these two main diffusive transports, the net charge transport (as represented by a diffusive flux of unit charges, $e\textsuperscript{+}$) was smaller than the transports of Na\textsuperscript{+} and K\textsuperscript{+} separately. However, diffusive transports still gave rise to a net electrical transport that was smaller, but of the same order of magnitude as the Ohmic current (compare fluxes of $e^+$ in Fig. \ref{Fflows}B1 and B2).

Fig. \ref{Fflows}B3 shows the total ionic transports ($j^f + j^d$) in the case (ii) when diffusion was included. As expected, the extracellular ionic transports differed quite significantly from the case when diffusion was not included (Fig. \ref{Fflows}A). However, the net transport of unit charges in the system were identical in the two cases (compare fluxes of $e+$ in Fig. \ref{Fflows}A with that in Fig. \ref{Fflows}B3). There is a simple physical argument to why this had to be the case: Since the formalism was based on Kirchoff's law (the sum of electrical currents into any given ECS volume were zero), the transmembrane current sources into/out from an extracellular volume had to be balanced by extracellular currents out from/into the same volume. As our simulations were set up so that the neurodynamics (and thus the transmembrane current sources) were identical in the cases with and without diffusion, also the net extracellular current had to be identical.

\subsection*{Effect of diffusion on extracellular potentials at shorter time scales}
Above, we predicted that realistic concentration variations and the diffusive currents that they evoked could cause observable variations in sustained extracellular voltage profiles, as earlier discussed in \cite{Dietzel1989}. As we saw in Fig. \ref{Fflows}, diffusive charge transport out of the soma region was of the same magnitude as field driven charge transport (compare fluxes of $e^+$ in Fig. \ref{Fflows}B1 and B2). The assumption that diffusive currents is negligible compared to field currents is this not warranted in the scenarios studied here, which had large, but realistic extracellular ion concentration gradients.

\begin{figure}[!ht]
\begin{center}
\includegraphics[width=6in]{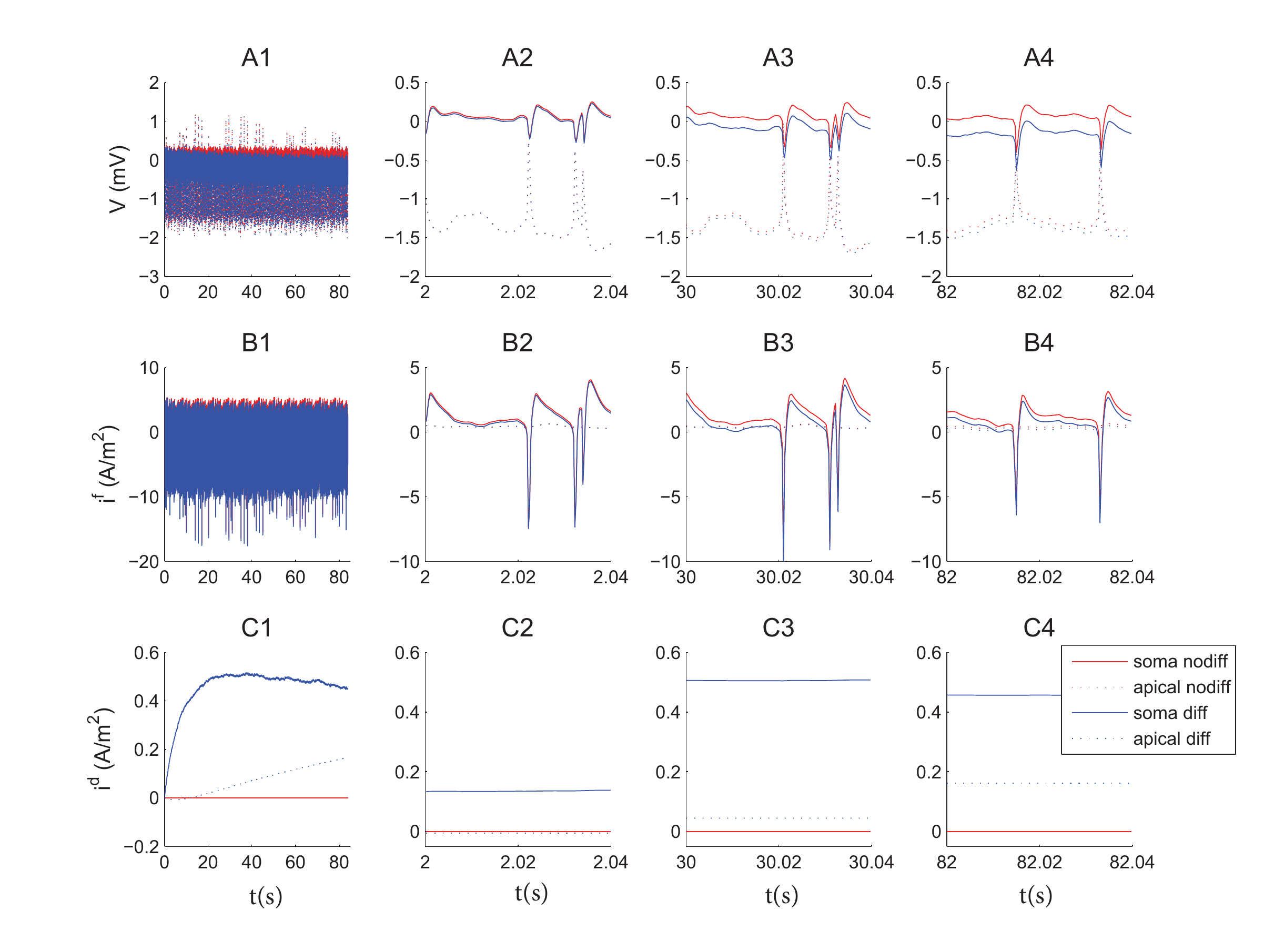}
\end{center}
\caption{\textbf{Dynamics on a shorter time-scale}. The time development of \textbf{(A)} the extracellular potential, \textbf{(B)} the field current, and \textbf{(C)} the diffusive current in ECS volumes surrounding the soma ($n=3$, full lines) and apical dendrite ($n=13$, dotted lines). The first column (A1-C1) shows the signal for the entire 84s simulation, while the remaining panels show the signal in selected, brief intervals during the simulations. Red lines show the signal obtained when diffusion was assumed to be zero, while blue lines show the signal obtained with the full electrodiffusive formalism. Field currents varied at the same time-scale as $V$ (ms), while diffusive currents varied very slowly (s).
}
\label{Ftemporal}
\end{figure}

The key differences between diffusive and field driven currents are clear if we explore their effects at a higher temporal resolution. In Fig. \ref{Ftemporal}A1 we have plotted the extracellular voltage at two selected depth intervals (soma $n=3$ and apical dendrite $n=13$), in the cases (i) without and (ii) with diffusion included. The figure shows how $V$ varies over the entire simulation of 84 s. It was mainly included to illustrate the general time course of $V$ during the simulation. In Fig. \ref{Ftemporal}B1 we see that the fluctuations in the field current (in the positive $z$-direction) out from the soma and apical dendrites had similar time course as $V$. As Fig. \ref{Ftemporal}C1 shows, this was definitely not the case for the diffusive current. While field current densities could vary by $\sim 10 A/m^2$ over the time course of a millisecond, it took about 30s of neural activity to build up an extracellular diffusive current density of $\sim 0.5 A/m^2$ (Fig. \ref{Ftemporal}C1). Early in the simulation, the diffusive current out from the soma-region increased with time in an approximately linear fashion. With time, diffusion between neighbouring subvolumes acted to smoothen the ECS ion concentration gradients. Due to this flattening of extracellular concentration gradients, the diffusive current out from the soma peaked after around 30 s, and decreased slowly after that. The concentration build-up was slower in the dendritic region, and the diffusive current out of the apical dendrite still increased in a linear fashion at the end of the 84 s simulation \ref{Ftemporal}C1.

Fig. \ref{Ftemporal}A2-4 shows the time course of $V$ at selected, shorter (40 ms) time intervals, which include include a few neuronal APs. In the soma-region, APs caused an initial decrease in the ECS voltage (current into the neurons discharge the ECS), followed by an increase in the ECS voltage when the neuron repolarized. Since currents always form closed loops, an inward current in the soma region evoked outward currents along the dendritic branches. Therefore, AP-signatures in the apical ECS region had the opposite temporal profiles (increase followed by decrease) compared to what we observed in the soma-region (decrease followed by increase). Although the fluctuation in $V$ during APs were of the same order of magnitude in the soma- and apical region, field currents were generally largest in the soma region (or out from the soma-region). The explanation to this has to do with the opposite extracellular AP-profiles in the subvolumes containing the soma versus those containing dendrites. In neighboring dendritic subvolumes, the AP profiles were very similar. Accordingly, the difference in $V$ between the subvolumes were small, and gave rise to small electrical currents through the ECS. Conversely, the soma-region had AP profiles that were opposite from its neighbours. During APs, there were thus large voltage differences between the soma ($n=3$) and its neighboring subvolumes, leading to to strong field currents. These observations are in line with previous investigations of extracellular AP-signatures (see e.g., \cite{Pettersen2008}).

At an early stage in the simulation, when ion concentrations did not diverge very much from the basal concentrations, the time development of $V$ was almost identical in the cases without (i) and with (ii) diffusion (Fig. \ref{Ftemporal}A2). However, as ion concentration built up in the system, $V$ was gradually shifted to more negative values in the case when diffusion was included (Fig. \ref{Ftemporal}A3-A4). After 82 s, $V$ was shifted with about -0.2 mV in the case when diffusion was included compared to the case with only field currents, consistent with what we also saw in Fig.\ref{Fconc}. These shifts took place at a slow time scale, so that $V$ was close to parallel in the diffusive and non-diffusive cases over the short 40 ms time intervals plotted in Fig. \ref{Ftemporal}A2-A4. Related to the shifts in $V$, also the field currents showed closely parallel time courses in the diffusive and non-diffusive cases (Fig. \ref{Ftemporal}B2-B4). At this short time courses, the diffusive currents remained roughly constant, and single APs evoked no visible fluctuations in diffusive currents (Fig. \ref{Ftemporal}C2-C4).

As we have seen, the maximum magnitude of the field currents were much larger than the diffusive currents during dramatic events such as APs (compare Fig. \ref{Ftemporal}B2-B4 with Fig. \ref{Ftemporal}C2-C4). In addition, we saw that diffusive currents had no visible impact on the temporal development of brief extracellular voltage signals, and would e.g. not have any impact on AP shapes detected in extracellular recordings (such as multi-unit array recordings). However, the high amplitude field currents were brief in duration, and for most of the time-course (i.e. between APs), the magnitude of field currents and diffusive currents were similar. We therefore hypothesized that diffusive currents could give a significant contribution to the low frequency part of extracellular voltage recordings, which are often sampled down to frequencies of 0.1-0.2Hz (see e.g., \cite{Einevoll2007}). Below, we explore which frequency components of the extracellular potential that may be influenced by diffusive sources.

\subsection*{Effect of diffusive currents on power spectra}
\begin{figure}[!ht]
\begin{center}
\includegraphics[width=6in]{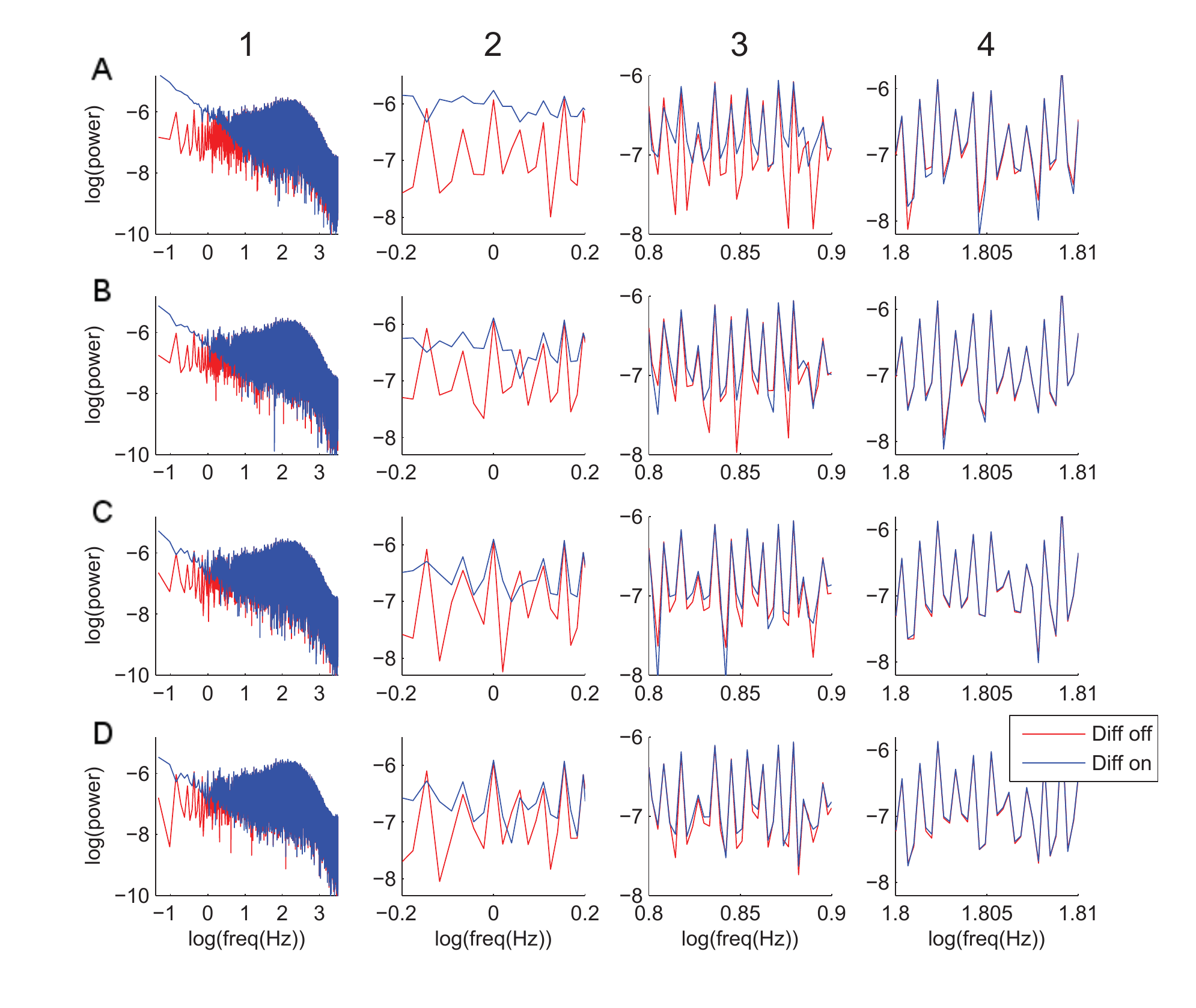}
\end{center}
\caption{\textbf{Power spectrum for the ECS potential in the soma region} Power spectra for $V$ in the case when diffusion was set to zero (red lines) and in the case when diffusion was included (blue lines). Rows \textbf{(A)-(D)} represent different 21s time intervals of the simulations, so that e.g. \textbf{(A)} shows the power spectrum for $V$ between $t=0$ and $t=21s$, and \textbf{(C)} is the power spectrum for $V$ between $t=42s$ and $t=63s$. Column 1 shows the full power spectrum, while columns 2-4 show close ups of selected frequency intervals. Diffusive currents had strongest impact in the first simulation interval (A), where it had a notable impact on the power spectrum up to frequencies as high as 7Hz (A3). The soma region had $n=3$.
}
\label{Fpowersoma}
\end{figure}

Fig. \ref{Fpowersoma} shows the power spectrum of $V$ at the depth level of the soma $n=3$ (i.e. of the signal seen in Fig. \ref{Ftemporal}A1). To see if the power spectrum varied over the time course of the simulation, we split the signal into four intervals of 21 s, which are shown in Fig. \ref{Fpowersoma}A, B, C and D.

The effect of diffusion on the power spectrum was largest in the first 21 s of the simulation, i.e. in the period when the diffusive current out from the soma increased steeply with time (Fig. \ref{Ftemporal}C1). Fig. \ref{Fpowersoma}A1 shows the full frequency spectrum for these initial 21 s, while Fig. \ref{Fpowersoma}A2-A4 are close-ups of selected frequency intervals of the signal in Fig. \ref{Fpowersoma}A1. When comparing the cases (i) without and (ii) with diffusion included, we found that diffusion had a strong impact on the power spectrum for frequencies around 1Hz \ref{Fpowersoma}A2, and even clearly influenced the power of frequencies almost up to 10 Hz (Fig. \ref{Fpowersoma}A3). For higher frequencies, the power spectra obtained without and with diffusion included were more or less identical (Fig. \ref{Fpowersoma}A4). In the final stages of the simulation (Fig. \ref{Fpowersoma}D), the range of frequencies that were influenced by diffusion terminated at lower frequencies. The power spectra obtained without and with diffusion closely coincided at 7Hz (Fig. \ref{Fpowersoma}D3), but diffusion still had a strong impact on frequencies up to 1 Hz (Fig. \ref{Fpowersoma}D2).

\begin{figure}[!ht]
\begin{center}
\includegraphics[width=6in]{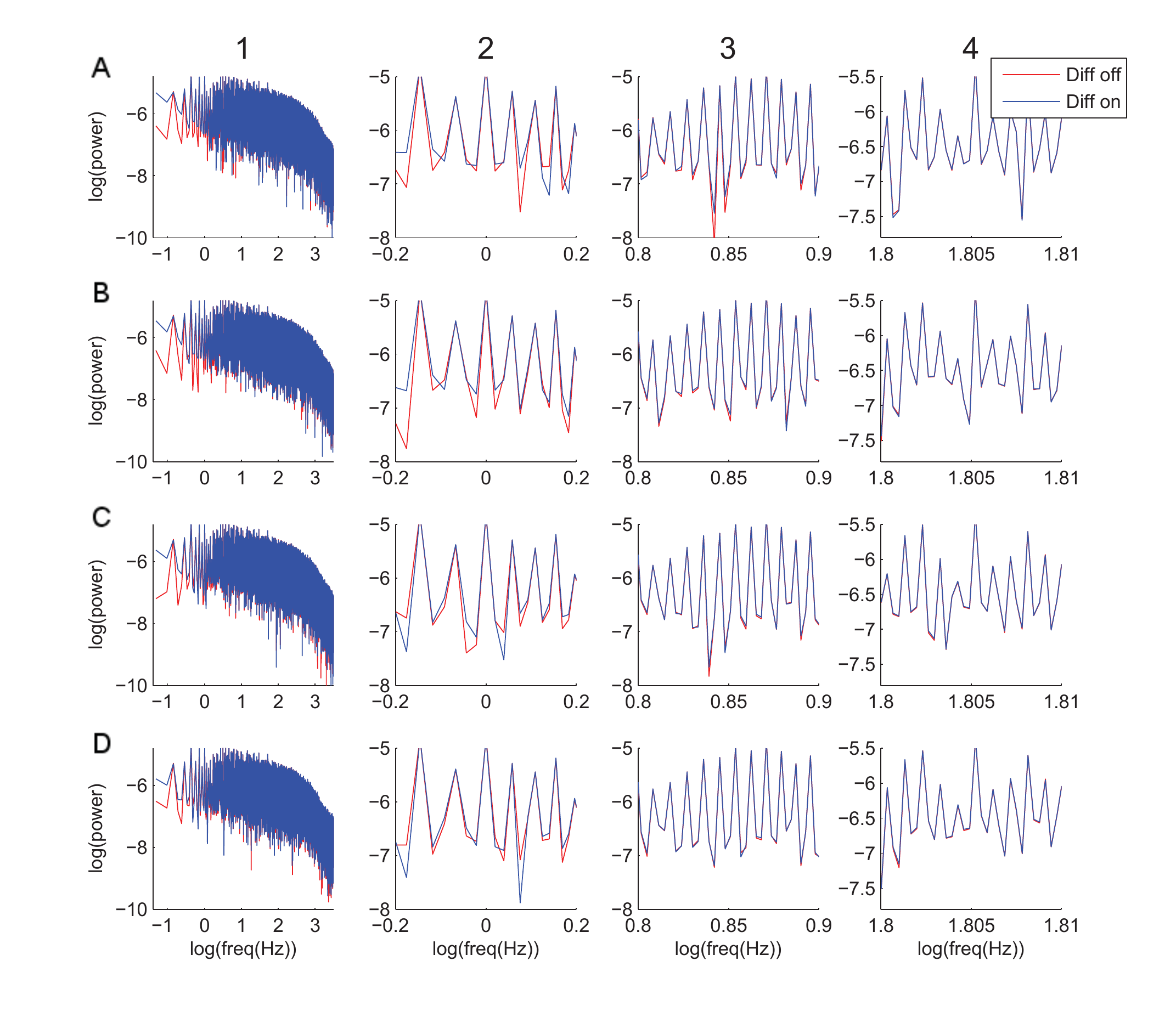}
\end{center}
\caption{\textbf{Power spectrum for the ECS potential in the apical region} Power spectra for $V$ in the case when diffusion was set to zero (red lines) and in the case when diffusion was included (blue lines). Rows \textbf{(A)-(D)} represent different 21s time intervals of the simulations, so that e.g. \textbf{(A)} shows the power spectrum for $V$ between $t=0$ and $t=21s$, and \textbf{(C)} is the power spectrum for $V$ between $t=42s$ and $t=63s$. Column 1 shows the full power spectrum, while columns 2-4 show close ups of selected frequency intervals. Diffusive currents had relatively small effects on the frequency spectrum in the apical region, since concentration changes were quite small here. The apical region considered here, had $n=13$.
}
\label{Fpowerapical}
\end{figure}

Fig. \ref{Fpowerapical} shows the frequency spectra for the extracellular potential in the region of the apical dendrites $n=13$. In this region, extracellular ion concentrations built up very slowly, and diffusion had a smaller impact. Here, only the very low frequency part (0.1 Hz) of the power spectrum was significantly influenced by including diffusion in the ECS dynamics.

\subsection*{Diffusive currents without neuronal sources}
A major component of the ECS diffusive currents did not depend directly on transmembrane neuronal sources. To show this, we ran an additional simulation, where we removed the neuronal sources midways in the simulation, i.e. after 42s. After this, extracellular transports were solely evoked by the concentration gradients that had built up in the ECS during the 42s of neuronal activity. In the case when diffusion was not included in the system, the ECS voltage gradient instantly turned to zero when the neuronal sources were removed, and there were no extracellular transports (results not shown).

\begin{figure}[!ht]
\begin{center}
\includegraphics[width=6in]{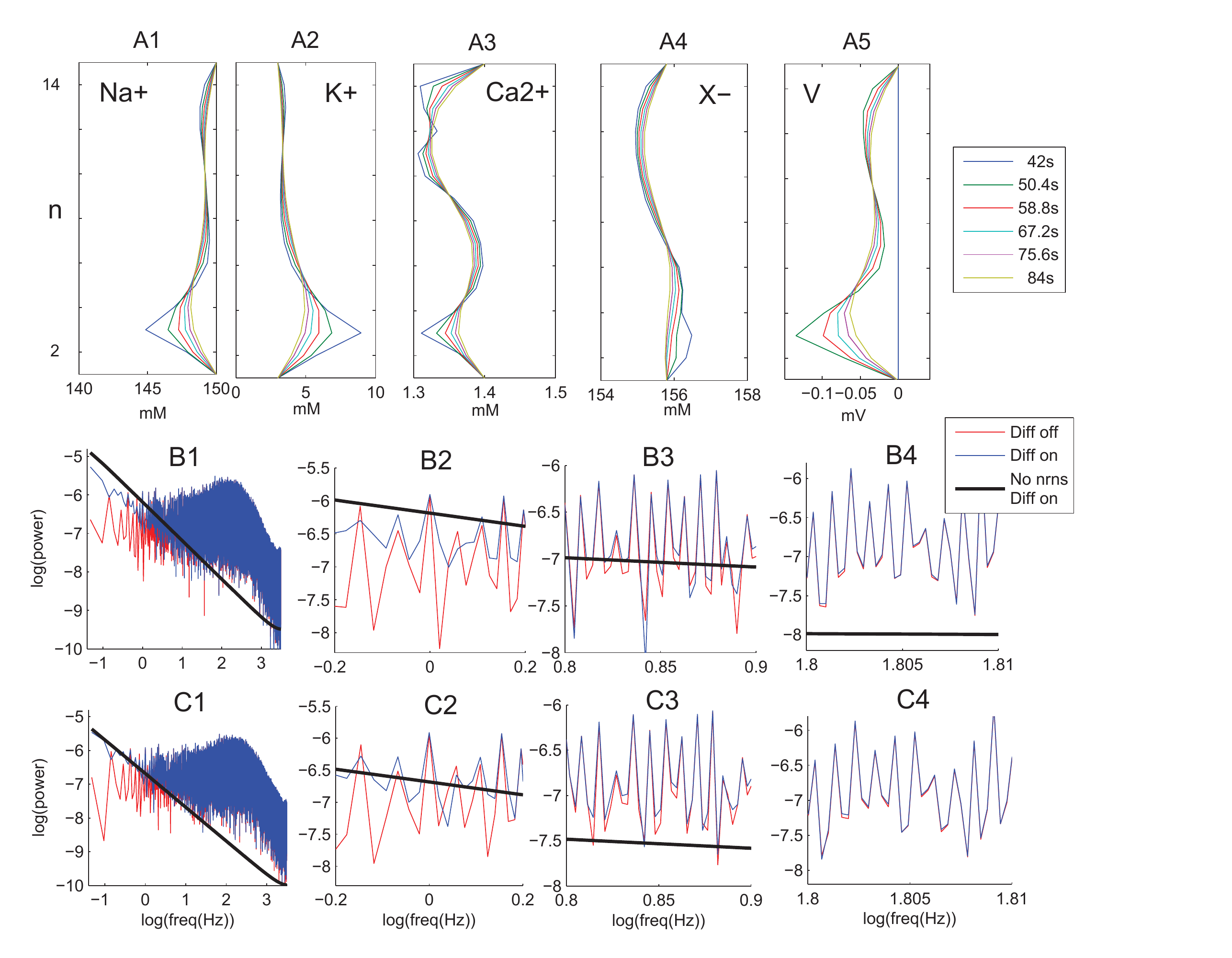}
\end{center}
\caption{\textbf{Extracellular dynamics in the case of no neuronal sources.} The neuronal transmembrane sources were removed midways (after 42s) in the (84s) simulation. \textbf{(A)} shows the distribution of different system variables over the depth of the cortex for different time points after the neurons were removed. Extracellular ion concentration gradients (\textbf{(A1)-(A4)} evoked diffusive currents that gave rise to liquid junction potentials in the ECS (\textbf{(A5)}). $\langle V \rangle$ was the temporal mean taken over the 8.4s prior to the value indicated in the legend. \textbf{(B-C)} Power spectra for $V$ in the time window $t=42s$ to $t=63s$  \textbf{(B)} and for $t=63s$ to $t = 84s$ \textbf{(C)}. \textbf{(B1) and \textbf{(C1)}} show the full power spectrum, while \textbf{B2-B4)} and \textbf{(C2)-(C4)} show close-ups of selected frequency intervals. Black, straight lines show the power spectra for the simulation when neuronal transmembrane sources were removed, and indicate an exponential decay of $\langle V \rangle$. Cases when neuronal sources were included (same as previous figures: blue lines = diffusion, red line = no diffusion) were included for comparison. Power spectra were for the soma region ($n=3$).
}
\label{Fsilence}
\end{figure}

In the case when diffusion was included in the system, the remaining concentration gradients evoked diffusive currents through the cortical column (Fig. \ref{Fsilence}). The diffusive currents evoked a non-zero voltage gradient through the column, so that the ECS potential in the soma region was about 0.1 mV more negative than the potential in the apical region. This diffusion-generated potential difference is known as the liquid junction potential (Fig. \ref{Fsilence}A5). To explain this effect, consider an example where two pools of salt solutions, so that pool 1 contains NaCl and pool 2 contains KCl. If these are set in contact with each other, Na\textsuperscript{+} will will diffuse from pool 1 to pool 2, while K\textsuperscript{+} will diffuse from pool 2 to pool 1. Since K\textsuperscript{+} and Na\textsuperscript{+} do not have identical diffusion constants ($D_K>D_{Na}$), diffusion will lead to net transport of charge from pool 2 to pool 1. It is known that the local charge separation associated with this process is extremely small \cite{Britz2014}, as charge accumulation is rapidly counteracted by an induced electrical potential difference between the two pools, which drives charge in the opposite direction from diffusion (c.f., the potential profile seen in Fig. \ref{Fsilence}). After the neuronal output was removed, the ionic composition was so that it evoked a net diffusive current out from the soma region. Accordingly, the electrical potential in the soma region decreased, which induced a net field current into the soma region. In this case, there was no neuronal current source, and the sum of the diffusive and field driven currents was always zero.

In the absence of neuronal sources, diffusive transports gradually reduced the ECS concentration gradients. This was a quite slow process, which happened over a time course of tens of seconds (Fig. \ref{Fsilence}A1-A4). The extracellular (liquid junction-) potential gradient decreased accordingly (Fig. \ref{Fsilence}A5). This gives us valuable insight in the diffusive processes, as the simulations in this case are solely dependent on the ECS concentration gradients, and does not depend on momentary concentration fluctuation generated by neuronal output, or by the neural processes that originally generated these concentration gradients.

In Fig. \ref{Fsilence}B-C we have explored the power spectrum of the diffusion evoked ECS potential (in the soma region). The black, straight lines indicate that the local $V$ decays exponentially with time. Fig. \ref{Fsilence}B shows the power spectrum for the time period between 42s and 63s, i.e. for the first 21s after the neuronal sources had been removed. When we compared the power-spectrum of the exponential decay with the previous power spectra obtained for this time period (i.e. when neuronal sources were present), we see that the removal of neuronal sources increased the power for frequencies up to about 7Hz. We also plotted the power spectrum for the following 21s of activity, i.e. for the time period between 63s and 84s (Fig. \ref{Fsilence}C). In this time interval, the ECS concentration gradients were smaller. The absence of neuronal sources then only increased the power for frequencies up to about 1Hz.

An increase in power for low frequencies was an expected outcome of removing the neuronal sources. In the presence of neuronal sources, the slow ECS potential profile was more or less preserved over time. The observed increase in power when neuronal sources were removed, reflected the gradual decay of the liquid junction potential $V$. However, we did not expect the range over which this decay process could dominate to stretch to as high frequencies as 1Hz or above.

Based on the analysis of Figures \ref{Fpowersoma}-\ref{Fsilence} we conclude that, as a generality, the contribution of diffusive sources to extracellular potentials are not negligible. However, a comparison between Fig. \ref{Fsilence}B and Fig. \ref{Fsilence}C, also shows that the powers of the extracellular $V$ that are affected by diffusive processes depend strongly on the ion concentration gradients in the system. Whether diffusive effects needs to be accounted for when interpreting extracellular recordings, thus depend on the extracellular ion concentration changes that are expected under the relevant experimental conditions (see Discussion for more on this).

\section*{Discussion}
We tested the hypothesis that neuronal activity could generate extracellular ion concentration gradients sufficiently large to induce diffusive currents of the same order of magnitude as field driven currents in the ECS. To explore this, we simulated the extracellular transport of ions in a cortical column, resulting from the activity of a small population of layer 5 pyramidal cells. We compared simulations when diffusive currents were included with simulations where diffusive currents were set to zero. Our findings were surprising. Not only could the slow component of diffusive currents be of roughly similar magnitude as field driven currents, but diffusive currents could influence the power spectrum of extracellular potentials up to frequencies as high as almost 10 Hz. We note that the simulated shifts in ECS ion concentrations were in the upper range of concentration shifts that have been observed under non-pathological, experimental conditions. Presumably, the role of diffusion was therefore in the upper range of what could be expected under physiological conditions, and it is possible that there are many cases when it is warranted to neglect diffusive currents (however, this should be verified in each specific case). We still conclude that, as a generality, diffusive currents can not be assumed to have a negligible impact on extracellular potentials, unlike what has been assumed in many previous theoretical analysis \cite{Riera2012, Pettersen2008, Reimann2013, Einevoll2007, Holt1999}.

\subsection*{Model limitations}
The simplified model setup used in the current study has several limitations. Firstly, cortical columns contain several neuron species, whose somata are located in different cortical layers. The small population of 10 layer 5 pyramidal cells used in the current study, will likely create a bias towards strong concentration gradients in layer 5 (or in the soma region $n=3$ in Fig. \ref{Fvox}). Secondly, synaptic connections between neurons were not considered in the current study. Such connections could induce a level of synchrony in the neuronal firing, which likely would have an impact on the power spectrum of the ECS potential \cite{Linden2011}. Thirdly, the multi-compartmental neuronal model used in this study \cite{Hay2011} (and most other multi-compartmental models) does not include ionic uptake mechanisms (such as Na\textsuperscript{+}/K\textsuperscript{+}-pumps). Such mechanisms would generally act to maintain the ECS ion concentrations closer to the basal levels, and also astrocytes are known to play a major role in the maintenance of the ECS \cite{Wang2008, Halnes2013}. As such mechanisms were not included in the current study, the simulated shifts in ion concentrations were most likely larger than those that would naturally occur from the neuronal activity. A natural way to improve the model would be to incorporate the effect of neuronal and glial ionic uptake-mechanisms, as well as more of the cortical complexity. If appropriately expanded in this way, the model could ideally replicate the exact relationship between neuronal activity and extracellular ion concentration dynamics, and could then be used to identify the exact experimental conditions under which diffusion is likely to have an impact on ECS potentials, and the conditions for which diffusive currents can rightfully be neglected.

In the following sub-sections we discuss some of these model limitations in further detail. We also argue that, despite its limitations, the modelling study presented here gives strong support to the conclusion that, as a generality, diffusive currents can not be assumed to have a negligible impact on extracellular potentials.

\subsection*{Interpretation of the extracellular potential}
In the current work, $V$ was computed in the following way: First, we determined what the current density $i^f$ needed to be between two ECS subvolumes in order for Kirchoff's current law to be fulfilled (Eq. \ref{FchargeB}), assuming that $i^f$ was uniformly distributed over the intersection ($A_c$) between the subvolumes. Next, we computed what $V$ needed to be in each subvolume in order to give the correct values for $i^f$ between subvolumes (c.f., Eq. \ref{Ifield}). The obtained values for $V$ thus represented average potentials over the entire ECS subvolumes (with volumes $l_c \times A_c$).

Experimentally, ECS potentials depend on the distances between the recording electrode and the neuronal current source \cite{Pettersen2006}. For example, ECS signatures of action potentials are much higher close to the neural membrane, while slower signals can have a longer spatial reach \cite{Pettersen2008, Linden2011}. Accordingly, it is likely that currents are not uniformly distributed over ECS cross sections $A_c$, but are larger in regions that are in the vicinity of neuronal membranes. A direct comparison between $V$ as determined by the formalism presented here (averaged over a ECS subvolume), and $V$ measured by point electrodes is thus not feasible. However, in both cases $V$ was determined by neuronal current sources, and followed the same time course as these signals (Fig. \ref{Ftemporal}). In addition, the estimates of $V$ in the current work showed sustained ECS profiles (Fig. \ref{Fconc}) that were qualitatively similar to those observed experimentally \cite{Dietzel1989, Cordingley1978}. We thus believe that the large scale (volume averaged) $V$ in the current study represents a useful variable for assessing relative contribution of field currents and diffusive currents at a tissue level.

In our model, we assumed that the average cortex surface area per neuron was about $300 {\mu}m^2$ \cite{Linden2011}. This number is species- and region specific (e.g. a smaller surface area of $125 {\mu}m^2$ per neuron was used in \cite{Pettersen2008}). Essentially, changing $A_c$ (while keeping the population size constant) would amount to changing the average distance between an arbitrary point in the ECS and the neuronal sources. The computed amplitudes in $V$ thus depended on the ECS cross section area $A_c$. However, the results regarding the relative contributions from diffusive versus field driven currents did not depend qualitatively on $A_c$. To see why, we can start by noting that the neuronal current sources at each depth layer $n$ were unaltered when the ECS volume was changed. Accordingly, the total ECS current $I^{tot} = i^{tot}A_c = (i^d + i^f)A_c$ in the $z$-direction would also be unaltered. An increase in $A_c$ by a factor $\alpha$, would reduce ECS ion concentration gradients by the same factor $\alpha$ (as the same number of ions would enter an enlarged volume, concentration changes would be smaller). This would lead to a reduction $i^d \rightarrow i^d/\alpha$ in the the diffusive current density, but at the same time an increase in the ECS cross section area over which diffusion occurs $A_c \rightarrow \alpha A_c$. The net diffusive current $I^d = i^dA_c$ would thus be invariant to changes in $A_c$. The same must therefore hold for the field current $I^f$. In analog to the situation with $i^d$, this would imply a reduction in the current density $i^f$ by a factor $\alpha$, caused by a decrease in the voltage gradients by the same factor $\alpha$. Thus, the amplitudes of ECS potentials would scale approximately linearly with $A_c$, but but the relative roles of diffusive versus field currents in generating these potentials would remain the same (the scaling is not strictly linear due to possible variations in the ECS conductivity, see further below). This was verified in additional simulations, where we varied $A_c$ (results not shown).

We also ran test simulations to explore how the results depended on size of the neuronal population (results not shown, but described here). If we used a population of 100 neurons instead of 10 neurons, and scaled up the ECS volume accordingly by a factor 10, the power of the $V$ signal was generally reduced (results not shown).  This was because the neurons were not firing in synchrony, and thus produced partially uncorrelated membrane currents within a given subvolume. The summation rules for correlated versus uncorrelated signals says that $N$ correlated signals, all with standard deviations (amplitudes) $\sigma$, will give a summed signal with standard deviation $\sigma N$, whereas $N$ uncorrelated signals with standard deviations $\sigma$ will give a summed signal with standard deviation $\sigma \sqrt(N)$. Thus, only fully correlated sources would give a linear increase in the net signal. A similar effect was shown in a previous study, where the high frequency part of the ECS potential was found to scale sublinearly with the number of APs elicited in a volume \cite{Pettersen2008}. Although changes in population size resulted in negative shifts in the power spectra, the qualitative findings regarding the relative contribution of diffusion to such power spectra were not significantly changed.

\subsection*{Extracellular conductivity}
Physically, the ECS conductivity ($\sigma$) is determined by the number of free charge carriers, weighted by their mobility and valence. Thus, in our simulations $\sigma$ was a varying function of the ECS concentrations (c.f., Eq. \ref{conductivity}). However, as absolute variations in ion concentrations were relatively small compared to the initial ion concentrations (at least of the most abundant species), and as these variations typically were asymmetric (decreases in K\textsuperscript{+} were accompanied by increases in Na\textsuperscript{+}), variations in $\sigma$ were only by a few percent compared to the initial value. By running test simulations where we kept $\sigma$ fixed at the initial value, we could confirm that these variations had no significant effect on our simulation results.

With the initial ion concentrations that we used, we obtained an ECS conductivity of $\sigma 0.76 S/m$. In the literature, there are quite some variations in values that are given for the ECS conductivity, and also variations in how this quantity is defined. In analysis of neural tissue, Chen and Nicholson \cite{Chen2000} operated with an apparent conductivity, defined as ${\sigma}' = \alpha/\lambda^2 \sigma$, which was used to describe extracellular electrical currents through neural tissue (as a whole) at a relatively large spatial scale. In that study, $\sigma$ represented the conventional conductivity associated with the extracellular bath solution. The extracellular tortuosity ($\lambda = 1.6$) represented the hindrances imposed on moving ions by e.g. neuronal processes, while $\alpha$ represented the fraction of the neural tissue that was actually ECS \cite{Nicholson1981, Nicholson1998}. The apparent ECS conductivity was in that work estimated to be 0.1 S/m \cite{Chen2000}. In our simulations, we considered the ECS as a separate domain, and thus explicitly accounted for the fact that ECS currents only moved through a fraction $\alpha = 0.2$ of the total tissue volume, but we did account for the hindrances (tortuosities) in the same way as Chen and Nicholson did. The apparent conductivity in the work by Chen and Nicholson \cite{Chen2000} would therefore correspond to a conductivity of $\sigma = 0.5 S/m$ in our ECS domain. For other comparisons, previous computational studies of local field potentials and current source density estimates, have used the value ${\sigma}' = 0.3 S/m$ \cite{Pettersen2008, Pettersen2008a, Linden2011}. However, also in those studies, ${\sigma}'$ represented the conductivity for electrical currents through neuronal tissue as a whole. This conductivity would therefore correspond to a conductivity of $\sigma = 1.5 S/m$ in our ECS domain. Our estimate for $\sigma$ thus lies between the previously estimated values for $\sigma$, and is relatively close to the value used in Chen and Nicholson's work \cite{Chen2000}.

\subsubsection*{Ion exchange between neurons and the ECS}
Commonly, only a subset of the transmembrane currents in a multi-compartmental neuron model are ion specific. In the model that we used \cite{Hay2011}, non-specific currents included the passive (leakage) current, synaptic currents, and the currents through a non-specific ion channel ($I_h$). For simplicity, we
assumed that all non-specified currents were carried by an unspecified ion species X\textsuperscript{-}, which to a large degree had the role that Cl\textsuperscript{-} has in the biological system. We are aware that this is an inaccurate assumption. However, we do not believe it to be of any significance for our main results. The arguments for this are (i) that ECS ion concentrations did not have any significant effects on the ECS conductivity (as discussed above), and (ii) that diffusive currents in the ECS were mainly mediated by K\textsuperscript{+} and Na\textsuperscript{+} gradients. The influence of the X\textsuperscript{-}-dynamics was therefore rather minor, and subdividing the currents into different ionic species would likely not have any significant impact on the results. The ECS also contains other ion species than the ones included in our simulation, such as phosphorous and magnesium. However, the concentrations of these species are typically low compared to those of the main charge carriers, so that the omittance of these is unlikely to be a concern with our model.

A more critical issue is that real neurons contain Na\textsuperscript{+}/K\textsuperscript{+}-exchangers. As these typically work at a slower pace than the mechanisms involved in fast time scale neurodynamics, they are typically not included in neuronal models, and were not included in the model that we used \cite{Hay2011}. In addition to neural ion pumps, also glial cells, and particularly astrocytes, are involved in the maintenance of the extracellular space \cite{Gardner-Medwin1983, Newman1993, Chen2000, Wang2008, Halnes2013}. Significant changes in ECS ion concentrations are therefore likely to occur only in cases when the neuronal activity level is too intense for such clearance mechanisms to keep up. We are therefore aware that the ionic concentration gradients like those we predicted in Fig. \ref{Fconc} are likely to be an overestimation of the ion concentration gradients that would realistically build up during the relatively moderate AP-firing activity of the neuronal population applied in our model. However, the key conclusions regarding the contribution of diffusive currents depended on ion concentration gradients in the ECS, but not directly on the neuronal activity responsible for building up such gradients. The ion concentration changes that occurred during our simulations were in the range of experimentally observed values during non-pathological conditions \cite{Hertz2013, Chen2000, Newman1993, Haj-Yasein2011}. We therefore believe that our key conclusions are qualitatively sound.

The exact shape of the $\langle V \rangle$-profiles (Fig. \ref{Fconc}) depended strongly on how the synaptic input was distributed over the cortical depth. In all simulations (Figs. \ref{Fdata}-\ref{Fsilence}) synapses were distributed uniformly over the neuronal membrane. We did run some test simulations to explore if the main conclusions depended critically on this modelling choice (results not shown, but described here). Profiles similar to those in Fig. \ref{Fconc} were obtained when only the apical dendrites contained synapses. In the case when synapses were found exclusively in basal dendrites and soma, $\langle V \rangle$-profiles increased gradually from 0mV in the soma and basal dendrites to +2mV in the most superficial layers, and were almost a mirror image from what we observed with the uniform synapse distribution. Accordingly, the direction of electrical transports were reversed. However, all cases gave rise to qualitatively similar conclusion regarding the relative importance of extracellular diffusive currents. We therefore only included the scenario with a uniform synapse distribution in our main results.

The main objective of this work was to explore the effect of diffusion on the dynamics of the ECS potential. We therefore compared simulations obtained when we set $j^{dk} = 0$ (i.e. the case with purely Ohmic extracellular currents), with simulations where the extracellular ion concentration- and voltage dynamics were derived using the full electrodiffusive scheme. In the modelling setup, we assumed that there was no feedback between the ECS to the neurons. That is, we did not account for changes in neural reversal potentials due to changes in ECS ion concentrations \cite{Oyehaug2012, Park2006}, or ephaptic effects of ECS potentials on neuronal membrane potentials \cite{Bokil2001, Mori2008, Frohlich2010, Agudelo-toro2013}. This simplification gave us the advantage that we could have exactly the same neurodynamics when comparing the ECS dynamics in the cases without or with diffusion included in the ECS. Hence, we could be sure that the observed differences between the two scenarios were due to extracellular diffusion, and not indirect effects stemming from altered neurodynamics. In a more realistic scenario, the shifts in ECS ion concentrations that occurred during the time course of the simulation would induce changes in the neuronal firing patterns. Most likely, the increases in K\textsuperscript{+} in the ECS would make neurons more excitable, and increase the AP firing rate. However, when it comes to the conclusions regarding the relative contribution of diffusive versus field currents for a given underlying neurodynamics, we do not believe that the inclusion of feedback from the ECS dynamics to the neurodynamics would induce any qualitative changes in our findings.

Theoretically, the sum of currents through a closed surface, such as a neuron, should be zero. In Fig. \ref{Fdata}D6, we saw that the total sum of transmembrane neural currents was small but not zero. This was a numerical inaccuracy that could be improved by using smaller time steps in the Neuron-simulation. However, a non-zero input current is, however, not inconsistent with the ECS-formalism presented in this work. In principle, there is no demand that the selected ECS-volume should not contain partial membranes, such as e.g., a neural axon entering from a neuron located outside the column selected here. The formalism was based on Kirchoff's current law, but had a leaky boundary at $n=1$. The small net current entering/leaving the ECS from the neuronal sources (Fig. \ref{Fdata}D6) gave rise to a corresponding small current leaving/entering the system at $n=1$ (results not shown). The boundary condition at $n=N$ ensured that no net current could leave the system there.

\subsection*{A novel mathematical framework for simulating electrodiffusion in neural tissue}
The mathematical framework presented here represents a novel, general framework that can be used to compute the dynamics of ion concentrations and the electrical potential in the ECS surrounding multi-compartmental neuronal models or networks of such (such as e.g. the Blue-Brain-Simulator \cite{Markram2006}). The framework was particularly adapted to study dynamics at a large spatiotemporal scale (for active populations on neurons at the level of neural tissue). For this purpose, the framework is significantly more computationally efficient than other electrodiffusive frameworks based on the Poisson-Nernst-Planck equations \cite{Leonetti1998, Lu2007, Lopreore2008, Nanninga2008, Pods2013}. A future ambition is to expand this framework so that it (i) can account for neuronal and glial ionic uptake mechanisms (ion pumps), and (ii) can be used to simulate general, three dimensional transport processes in neural tissue. Such a generalized framework would be of undisputable value for the field of neuroscience, as it can be applied to explore pathological conditions related to ion concentration shifts in neural tissue \cite{Sykova2008, Enger2015, Frohlich2008, Florence2009}, and for further exporations of possible effects that diffusive currents can have on recorded extracellular fields.

\section*{Methods}

\subsection*{Extracellular dynamics}
From a methods-development point of view, the main contribution of the current work was the formulation of a novel formalism for computing the ion concentration dynamics and the electrical potential in the ECS surrounding a neural population. For simplicity, we assumed that spatial variation only occurred in one spatial direction ($z$-direction), and that we had radial homogeneity. This simplification may be warranted in several brain regions, such as within a cortical column.

\subsubsection*{Continuity equation}
The formalism represents a way of solving the continuity equation for the ionic concentrations ($c^k_n$ (mol/m\textsuperscript{3})) in a system as sketched in Fig. \ref{Fvox}B. The ECS is subdivided into a number of $N$ subvolumes of length $l_c$ and cross section area $A_c$ (in the application used in the current work, we had $N=15$). In each subvolume $n$, the concentrations of all present ion species $k$ are assumed to be known. Ions may enter the subvolume either via (i) transmembrane fluxes from neurons that exchange ions with the subvolume ($j^{kM}$), (ii) diffusive fluxes between neighboring subvolumes ($j^{kd}$), or (iii) field fluxes between neighboring subvolumes ($j^{kf}$). The formalism computes the ECS fluxes, and is general to the choice of neuronal sources. For now, we therefore assume that the transmembrane fluxes $j^{kM}$ for all ion species as well as the transmembrane capacitive current (which will be relevant below) are known (e.g., determined from a separate simulation using the e.g., the NEURON simulator). The continuity equation is:

\begin{equation}
A_{c}l_{c}\frac{\partial{c^{k}_n}}{\partial t} = J^{kM}_n + J^{kd}_{n-1,n}  - J^{kd}_{n,n+1} + J^{kf}_{n-1,n} - J^{kf}_{n,n+1}
\label{Fcontinuity}
\end{equation}
\noindent
Here, $A_{c}l_{c}$ is the volume of a subvolume, so that the left hand side of Eq. \ref{Fcontinuity} represents the time dependent change of the number of particles (in mol/s) of species $k$ in subvolume $n$.  The extracellular fluxes are described by the Nernst-Planck equation:

\begin{equation}
J^{kd}_{n-1,n} = -\frac{A_{c} D^{k}}{l_{c}} (c^k_n) - c^k_{n-1}) \\
\label{Jdiff}
\end{equation}
\noindent
and
\begin{equation}
J^{kf}_{n-1,n} = - \frac{A_{c} D^{k} z^k}{\psi l_{c}}  \frac{c^k_{n-1}+c^k_n}{2} (V_n-V_{n-1}),
\label{Jfield}
\end{equation}
\noindent
We have used the notation that $J_{n-1,n}$ denotes the flux from subvolume $n-1$ to subvolume $n$. The factor $\psi = RT/F$ is defined in terms of the gas constant ($R = 8.3144621 \, \mathrm{J/(mol \, K)}$), the absolute temperature ($T$), and Faraday's constant. Furthermore, $D^k = \tilde{D}^k/\lambda^2$ is the effective diffusion constant for ion species $k$, where $\tilde{D^k}$ is the diffusion constant for ion species $k$ in dilute solvents, and $\lambda$ is the extracellular tortuosity, which represents miscellaneous hindrances to motion through neuronal tissue \cite{Nicholson1998, Chen2000}. In the current work, we used the standard values \cite{Lyshevski2007}: $\tilde{D^K} = 1.96\times 10^{-9} m^2/s$, $\tilde{D^{Na}} = 1.33\times 10^{-9} m^2/s$, $\tilde{D^{Ca}}$ = $0.71\times 10^{-9} m^2/s$ and $\tilde{D^{X}} = 2.03\times 10^{-9} m^2/s$ (for the unspecified ion species, we used the diffusion constant for Cl\textsuperscript{-}). These values were modified with a tortuosity of $\lambda = 1.6$ \cite{Chen2000}.

We assume that the edge subvolumes ($n=1$ and $n=N$) represent a background where ion concentrations remain constant. The continuity equation then governs the ion concentration dynamics in all the $N-2$ interior subvolumes. If we include a number $K$ of different ion species, the continuity equation (Eq. \ref{Fcontinuity}) for $n = 2,3,...,N-1$ and $k = 1,2,...,K$ gives us $K(N-2)$ conditions for the $K(N-2)$ ion concentrations $c^{k}_n$ in the $N-2$ non-constant subvolumes. However, the continuity equation also includes $N$ state variables for $V_n$ in all subvolumes (including the edges). We thus need $N$ additional constraints to fully specify the system.

\subsubsection*{Derivation of the extracellular potential}
In the following, we derive expressions for the ECS potential ($V_n$) based on the principle of Kirchoff's current law, and the assumption that the bulk solution is electroneutral \cite{Halnes2013}. To do this, we multiply the continuity equation (Eq. \ref{Fcontinuity}) by $Fz^k$, take the sum over all ion species $k$, and obtain the continuity equation for electrical charge:
\begin{equation}
\frac{\partial{q_n}}{\partial t} = I^M_n + I^{d}_{n-1,1} - I^{d}_{n,n+1} + I^{f}_{n-1,n} - I^{f}_{n,n+1}
\label{Fcharge}
\end{equation}
Here, we have transformed fluxes/concentrations into electrical currents/charge densities by use of the relations:
\begin{equation}
I^M_n = F\sum_k \left(z^k J^{kM}_n \right),
\label{IMdef1}
\end{equation}
\noindent
\begin{equation}
q_n/(A_{c}l_{c}) = \rho_n = F\Sigma_k \left(z^k c^k_n \right),
\label{rhodef1}
\end{equation}
\noindent
\begin{equation}
I^{d}_{n-1,n} = F\Sigma_k \left(z^k J^{kd}_{n-1,n} \right) =
-A_{c} F\Sigma_k \left(\frac{z^k D^{k}}{l_{c}} (c^k_n-c^k_{n-1})) \right) \\
\label{Idiff}
\end{equation}
\noindent
and
\begin{equation}
I^{f}_{n-1,n} = F\sum_k \left(z^k J^{kf}_{n-1,n} \right) =
-\frac{A_{c}}{l_{c}}\sigma_{n-1,n}(V_n-V_{n-1}),
\label{Ifield}
\end{equation}
\noindent
where $z^k$ is the valence of ion species $k$ and $F=96,485.3365 \, \mathrm{C/mol}$ is Faraday's constant. In Eq.\ref{Ifield}, we also defined the conductivity (units $(\Omega m)^{-1}$)for currents between two subvolumes $n-1$ and $n$ as:
\begin{equation}
\sigma(n-1,n) = F\sum_k \left(\frac{D^{k} (z^k)^2}{\psi}\frac{c^k_{n-1}+c^k_n}{2}\right)
\label{conductivity}
\end{equation}
\noindent

At time scales of $>1ns$, bulk solutions can be assumed to be electroneutral \cite{Grodzinsky2011}. For our purpose, bulk electroneutrality implies that any net ionic charge entering an ECS-subvolume, must be identical to the charge that enters a capacitive neural membrane within this subvolume. This is also an implicit assumption in the cable equation, upon which the Neuron-simulator is based (see e.g., \cite{Rall1977, Koch1999, Qian1989, Halnes2013}). With this assumption at hand, the continuity equation for charge (Eq. \ref{Fcharge}) becomes useful for us, as it is governed by a constraint that we did not have at the level of ion concentrations (Eq. \ref{Fcontinuity}). Electroneutrality in the bulk solution implies that the net charge entering an ECS subvolume (the time derivative of $q_n$ in Eq. \ref{Fcharge}) must be identical to the charge which accumulates at the neuronal membrane and gives rise to the neuronynamics. This means that the time derivative of $q_n$ must be equal to the capacitive current that we know from the NEURON-simulator:
\begin{equation}
\frac{\partial{q_n}}{\partial t} = -I^{cap}_n
\label{Ficap}
\end{equation}
\noindent
Thus, $q_n$ (in Eq. \ref{Fcharge}) is not a state variable, but an entity known from the Neuron-simulation (i.e. an input condition to the ECS). With this at hand, we can rewrite Eq. \ref{Fcharge})on the form:
\begin{equation}
-I^{cap}_n - I^M_n = I^{d}_{n-1,1} - I^{d}_{n,n+1} + I^{f}_{n-1,n} - I^{f}_{n,n+1}
\label{FchargeB}
\end{equation}
\noindent
We now see that Eq. \ref{FchargeB} is simply Kirchoff's current law, and states that the net current into an ECS volume $n$ is zero, cf. Fig. \ref{Fvox}C. If we insert Eq. \ref{Ifield} for $I^f$, Eq. \ref{FchargeB} becomes:
\begin{equation}
\sigma_{n-1,n}V_{n-1} - (\sigma_{n-1,n} + \sigma_{n,n+1})V_n + \sigma_{n,n+1}V_{n+1}
= \frac{l_{c}}{A_{c}} \left(-I^{cap}_n - I^M_n - I^{d}_{n-1,1} + I^{d}_{n,n+1}\right)
\label{Fcharge3}
\end{equation}

We note that $I^M_n$ was defined as the net \emph{ionic} transmembrane current (Eq. \ref{IMdef1}), and that it does not include the capacitive current. We further note that Eq. \ref{Fcharge3} for a subvolume ($n$) depends on the voltage levels in the two neighbouring subvolumes ($n-1$ and $n+1$), and thus only gives us $N-2$ conditions, i.e. one for the $N-2$ inferior volumes. We need two additional criteria for the edge subvolumes ($n=1$ and $n=N$). As we may chose an arbitrary reference point for the voltage, we may take the first criterion to be:
\begin{equation}
V_1 = 0
\label{Fcharge0}
\end{equation}

As the second criterion, we impose a boundary condition stating that no net electrical current is allowed to pass between the subvolumes $n=N-1$ and $n=N$ (i.e. no net electrical current enters/leaves the system from/to the constant background). Since there may be a diffusive current between these two subvolumes ($c^k_{N-1}$ is not constant), this criterion implies that we must define $V_N$ so that the field current is opposite from the diffusive current ($I^{d}_{N-1,N} + I^{f}_{N-1,N} = 0$). If we insert for $I^f$ (c.f., Eq. \ref{Ifield}), this condition becomes:
\begin{equation}
\sigma_{N-1,N}V_{N-1} - \sigma_{N-1,N}V_N = \frac{l_{c}}{A_{c}} I^{d}_{N-1,1}
\label{FchargeN}
\end{equation}

The conductivities ($\sigma$) and the diffusive currents ($I^d$) are defined by ionic concentrations in the ECS, whereas we assumed that the neuronal output ($I^{cap}$ and $I^M$) was known. Equations \ref{Fcharge3}-\ref{FchargeN} thus give us $N$ equations for the $N$ voltage variables $V_n$. In matrix form, we can write the system of equations (Eq. \ref{Fcharge3}-\ref{FchargeN}) as:
\begin{equation}
A_{m,n}V_n = M_n,
\label{Vmatrix}
\end{equation}
where the vector $M_n$ has elements:
\begin{equation}
M_n =
\begin{cases}
0 &  \mathrm{for} \, n = 1 \\
\frac{l_{c}}{A_{c}} \left(-I^{cap}_n - I^M_n - I^{d}_{n-1,1} + I^{d}_{n,n+1}\right) &  \mathrm{for } \, n = 1, 2, ..., N-1 \\
\frac{l_{c}}{A_{c}} \left( I^{d}_{N-1,1} \right) &  \mathrm{for } \, n = N \\
\end{cases}
\label{Mn}
\end{equation}
\noindent
and where
\begin{equation}
A_{m,n} =
 \begin{pmatrix}
  A_{1,1} & A_{1,2} & 0 & 0 & \cdots & 0\\
  A_{2,1} & A_{2,2} & A_{2,3} & 0  & \cdots & 0\\
  0 & A_{3,2} & A_{3,3} & A_{3,4} & \cdots & 0 \\
  \vdots  & \ddots  & \ddots & \ddots & \ddots &\vdots \\
  0 & 0 & \cdots & A_{N-1, N-2} & A_{N-1, N-1} & A_{N-1,N} \\
  0 & 0 & \cdots & 0 & A_{N, N-1} & A_{N,N} \\
 \end{pmatrix}
 \label{Amatrix}
\end{equation}
\noindent
is a tridiagonal matrix. The diagonal above the main diagonal is given by:
\begin{equation}
A_{n,n+1} =
\begin{cases}
0 &  \mathrm{for} \, n = 1 \\
\sigma_{n,n+1} &  \mathrm{for} \, n = 2,3,...,N-1 \\
\end{cases}
\end{equation}
\noindent
The diagonal below the main diagonal is given by:
\begin{equation}
A_{n,n-1} =
\begin{cases}
\sigma_{n,n+1} &  \mathrm{for } \, n = 2,3,...,N \\
\end{cases}
\end{equation}
The main diagonal is given by:
\begin{equation}
A_{n,n} =
\begin{cases}
1 & \mathrm{for } \, n = 1 \\
- \left( \sigma(n-1,n) + \sigma(n,n+1) \right) &  \mathrm{for } \, n = 2,3,...,N-1 \\
- \sigma(N-1,N) &  \mathrm{for } \, n = N \\
\end{cases}
\end{equation}

For each time step in the simulation, we can determine $V_n$ by solving the algebraic equation set:
\begin{equation}
V_n = A_{m,n}^{-1}M_n,
\label{Vsolve}
\end{equation}

When we ran simulations where diffusion was not included, $j^d$ was simply set to zero in the continuity equation (Eq. \ref{Fcontinuity}), and in the equation where the ECS potential is derived (Eq.\ref{Mn}).

\subsubsection*{Initial Conditions}
As initial conditions, we assumed that all ECS volumes were at potential $V_n = 0$. The initial ion concentrations were also identical in all ECS subvolumes. We used $c^{K0} = 3 mM$, $c^{Na0} = 150 mM$, $c^{Ca0} = 1.4 mM$. These ion concentrations are quite typical for the ECS solutions \cite{alzet}. To obtain an initial charge density of zero in the bulk solution ($\sum z^k c^{k0} = 0$), we computed that the initial concentration for the unspecified anion was $c^{X0} = 155.8 mM$. This value for $c^{X0}$ is close to typical ECS concentrations for Cl\textsuperscript{-}, and the unspecified ion X\textsuperscript{-} can be seen as essentially taking the role that Cl\textsuperscript{-} has in real systems.

\subsubsection*{Power spectrum analysis}
The power spectra (Figs. \ref{Fpowersoma}-\ref{Fsilence}) were computed with the fast Fourier-transform in MATLAB (http://se.mathworks.com/).

\subsection*{Neuronal population dynamics}
In the current work, we applied the electrodiffusive formalism presented above to predict the extracellular ion concentration dynamics and electrical potential surrounding a small population of 10 layer 5 pyramidal cells. The neural simulation used in this study was briefly introduced in the Results section, but is presented in more detail here.

\subsubsection*{Pyramidal cell model}
As neural model, we used the thick-tufted layer 5 pyramidal cell model by Hay et al. \cite{Hay2011}, which was implemented in the NEURON simulation environment \cite{Hines2009}.  The model was morphologically detailed (it had 196 sections, each of which was divided to 20 segments), and had an extension of slightly less than $1300 \mu m$ from the tip of the basal dendrite to the tip of the apical dendrites. It contained ten active ion channels with different distributions over the somatodendritic membrane, including two Ca\textsuperscript{2+}-channels ($i^{CaT}$ and $i^{CaL}$), five K\textsuperscript{+}-channels ($i^{KT}$, $i^{KP}$, $i^{SK}$, $i^{Kv3.1}$, and $i^m$) and two Na\textsuperscript{2+}-channels ($i^{NaT}$ and $i^{NaS}$). In addition, in included a non-specific ion channel ($I^h$) and the non specific leakage current $i^{leak}$. We refer to the original publication for further model details \cite{Hay2011}.

\subsubsection*{Synapse model}
The neurons were driven by Poissonian input trains through 1000 synapses. The synapses were uniformly distributed across the membrane so that the expected number of synapses in a segment was proportional to its membrane area. A population of 10 neurons was simulated by running 10 independent simulations with the same neural model. In the independent simulations, we applied different input trains (but with the same Poissonian statistics).

The AMPA synapses were modelled as $\alpha$-shaped synaptic conductances:
\begin{equation}
I(t) = \left\{\begin{array}{l}g_{max}(t-t_0)/\tau\times\exp((t-t_0)/\tau),\ \mathrm{when}\ t \geq t_0\\0,\ \mathrm{when}\ t < t_0
\end{array}\right.,
\end{equation}
where $t_0$ represents the time of onset. The time constant was set to $\tau = 1.0$ms, and the synaptic weight $g_{max} = 0.0008686\mu$S was tuned so that the input evoked an average single-neuron AP firing rate of about 5 APs per second, which is a typical firing rate for cortical neurons \cite{Linden2011}.

\subsubsection*{Population output to the ECS}
The expansion of the cell morphology in the applied computational model \cite{Hay2011} was such that the maximal spatial distance between two segments (from tip of basal dendrite to tip of apical dendrite) was $<1300 \mu m$. We therefore assumed a cortical depth of $1500 {\mu}m$, and subdivided it into $N=15$ depth intervals of length $l_c = 100 {\mu}m$, so that the neurons populated the interior 13 subvolumes. Each neural segment was assigned as belonging to a particular subvolume $n$, determined by the spatial location of the segment midpoint. In the setup, the soma was placed in subvolume $n=3$, the basal dendrites were in subvolumes $n=2,3$ and 4, and the apical dendrites were in subvolumes $n=3,...,14$. The multi-compartmental model also included a short axon, which was, however, not based on the reconstruction and hence had no fixed coordinates. We assigned the axonal segments into the same subvolume as the soma, $n=3$. The edge-subvolumes 1 and 15 contained no neural segments (see Fig. \ref{Fvox}A).

The transmembrane current density ($i^{kM}_{seg}$) of ion species $k$ is available in the Neuron simulation environment. It was multiplied by the surface area of the segment ($A_{seg}$) to get the net current, and divided by Faraday's constant ($F$) to get a net ion flux with units mol/s. During the neural simulation, we grouped all currents that were carried by a specific ion species into the net transmembrane influx/efflux of this ion species. We assumed that all non-specific currents, including the synaptic currents ($i^{leak}$, $i^h$ and $i^{syn}$) were carried by a non specific anion that we denoted X\textsuperscript{-}. In this way we could compute the net efflux of each ion species into a subvolume $n$:

\begin{eqnarray}
J^{CaM}_n = \frac{1}{2F}\sum_{seg} \left((i^{CaT}_{seg} + i^{CaL}_{seg}) A_{seg} \right)\\
\nonumber
J^{NaM}_n = \frac{1}{F}\sum_{seg} \left((i^{NaT}_{seg} + i^{NaS}_{seg}) A_{seg} \right)\\
\nonumber
J^{KM}_n = \frac{1}{F}\sum_{seg} \left((i^{KT}_{seg} + i^{KP}_{seg} + i^{SK}_{seg} + i^{Kv3.1}_{seg} + i^m_{seg} )  A_{seg} \right)\\
\nonumber
J^{XM}_n = -\frac{1}{F}\sum_{seg} \left((i^{leak}_{seg} + i^h_{seg} + i^{syn}_{seg})  A_{seg} \right).
\label{JkM}
\end{eqnarray}
Here, the sum was taken over all neural segments ($seg$) of all 10 neurons contained in $n$. The factor 2 in the denominator in the extression for $J^{CaM}_n$ was due to $Ca^{2+}$ having valence 2, and the negative sign in the expression for $J^{XM}_n$ was due to $X^-$ having valence -1. We also kept track of the (non-ionic) capacitive currents, as required by the electrodiffusive formalism (Eq. \ref{Fcharge3}).

\begin{equation}
I^{cap}_n = \sum_{seg} \left( i^{cap}_{seg} A_{seg} \right)
\label{icapo}
\end{equation}

Following \cite{Linden2011}, we assumed that the average cortex surface area per neuron was about $300 \mu m^2$. As we had 10 neurons, and as only about 20 \% of cortical tissue is extracellular volume, the ECS subvolumes used in our simulations had surface area $A_c = 600 \mu m^2$ and length $l_c = 100 \mu m$ (Fig. \ref{Fvox}B).

We wanted to simulate the extracellular ion concentration dynamics for steady-state neuronal activity over 84 s. Due to the immense amount of data associated with recording all ionic currents in 13 depth layers over such a long time period, we only simulated the neurons for 7 s, and looped this data 12 times to obtain 84 s of neuronal output.

\bibliography{Arxiv_nvox}

\end{document}